\documentclass[aps,
showpacs,groupedaddress]{revtex4-1}

\def\bfpi{\mbox{\boldmath $\pi$}}
\newcommand{\be}{\begin{equation}}
\newcommand{\ee}{\end{equation}}
\newcommand{\nn}{\noindent}
\newcommand{\no}{\nonumber\\}
\newcommand{\ba}{\begin{eqnarray}}
\newcommand{\ea}{\end{eqnarray}}

\usepackage{graphicx}
\usepackage{amsfonts}
\usepackage{amssymb}
\begin{document}

\title{Parity Breaking Medium and Squeeze Operators
}


\author{A. A. Andrianov$^{a,b}$,
\footnote{Corresponding author: a.andrianov@spbu.ru}
S. S. Kolevatov$^{a}$\footnote{s.s.kolevatov@gmail.com}
and R. Soldati$^{c}$\footnote{ roberto.soldati@bo.infn.it}
}
\affiliation{$^a$ \, Saint-Petersburg State University, 7/9 Universitetskaya nab., St.Petersburg,
Russia}
\affiliation{$^b$\,
 Institut de Ci\`encies del Cosmos (ICCUB),
Universitat de Barcelona,
 Barcelona, Catalonia, Spain}
 \affiliation{$^c$\,
Dipartimento di Fisica e Astronomia - Universit\'{a} di Bologna and Istituto Nazionale di Fisica Nucleare - Sezione di Bologna, Italy}
\begin{abstract}
The transition between a Minkowski space region and a parity breaking medium domain is
thoroughly discussed. The requirement of continuity of the field operator content across the
separating boundary of the two domains leads to Bogolyubov transformations, squeezed pairs
states and squeeze operators that turn out to generate a functional SU(2) algebra.
According to this algebraic approach, the reflection and transmission probability amplitude
across the separating boundary are computed.
The probability rate of the emission or absorption of squeezed pairs out of the vacuum  (generalization of the  Sauter-Schwinger-Nikishov formula) is obtained.
\end{abstract}
\maketitle
\section{Introduction: possible physics of time- or space-dependent pseudoscalar condensate}
During the last decades the available bounds for validity of fundamental laws
in Physics have been attracting more and more
attention following succeeding experimental improvements,
both in the laboratory research and in astrophysics \cite{Carroll:1989vb}--\cite{Stecker:2009hj}.
More specifically, in Quantum Electrodynamics the interest towards possible Lorentz and CPT Invariance Violations (LIV for short) was raised up after the seminal paper \cite{Carroll:1989vb}, where the very possibility to deal with a parity odd vector background in the large scale Universe was  conjectured.
The latter was employed to modify QED by supplementing it with the Chern-Simons
(CS for short) term in the Action.
Later on the various aspects of its signatures were discussed
\cite{Andrianov:1994qv}--\cite{mauro}
although it has not yet been detected
\cite{Kostelecky:2002hh}--\cite{Cheng:2006us}.

In particular, spontaneous Lorentz symmetry breaking
may cause LIV after condensation of massless axion-like fields
\cite{Andrianov:1994qv},\cite{Kostelecky:2002hh}--\cite{ArkaniHamed:2004ar}.
Cold relic axions resulting from vacuum misalignment \cite{sikivie,raffelt}
in the early Universe
is a popular and so far viable candidate to dark matter. If we assume that
cold axions are the only contributors to the matter density of the Universe
apart from ordinary baryonic matter its density must be \cite{wmap} of the order
\be
\rho\,\simeq\, 10^{-30}\ {\rm g}\ {\rm cm}^{-3}\,\simeq\, 10^{-46}\ {\rm GeV}^4.
\ee
Of course dark matter is not uniformly distributed, its distribution
traces that of visible matter (or rather the other way round).
As well, on  stellar scales, the emergence of spontaneous Lorentz symmetry breaking in bubbles of pseudo-scalar condensates (of pions or axions)
may be detected in neutron stars making influence on their cooling rate \cite{aaek2016}.

Last decade several experiments in heavy ion collisions  have indicated an abnormal yield of lepton pairs of invariant mass $< 1$ GeV in the region of small rapidity and moderate transverse momenta
\cite{ceres,phenix,HADES} (reviewed in \cite{tserruya}).
This effect is visible only for collisions that are central or nearly central. Most studies refer to $e^+e^-$ pairs but dimuon pairs have also been found to be produced in excess above the dimuon threshold.
The explanation of this enhancement is outlined in \cite{aaep}
complementing the more conventional thermal effects. We conjecture that the effect may be a manifestation of
local parity breaking in colliding nuclei due to generation of pseudo-scalar, isosinglet or neutral isotriplet, classical
background whose magnitude and profile depends on the dynamics of the collision. Recently a possibility to generate an isotriplet pseudo-scalar condensate at large baryon densities has been argued for in \cite{anesp}. In \cite{kharzeev} it has been suggested that for peripheral interactions  a complementary effect should occur, namely,  an isosinglet pseudo-scalar background could appear
as the result of large-scale fluctuation of topological charge leading to the so-called Chiral Magnetic Effect (CME)
studied also by lattice QCD simulations \cite{lattice} and seemingly detected in the STAR experiments on RHIC \cite{star}.

All above mentioned phenomena are situated in  bounded volumes and a specific parity violating effect, namely, the gradient of isosinglet pseudo-scalar condensate can be formed near the volume boundary separating from a parity-even Maxwell QED vacuum. In the static situation this gradient is orthogonal to the boundary and it may generate spatial parity breaking in QED interacting to photons or more generally to vector mesons via the Chern-Simons interaction presumably induced by fermion polarization effects \cite{AACGS2010}.

In this paper we present a thorough analysis of the quantum theory of massive
vector fields on a flat space-time in the presence of a boundary. Specifically,
the boundary corresponds to a three-dimensional hyper-plane which separates two half-spaces
of the Minkowski space, on one side of which a Chern-Simons interaction is active.
We restrict ourselves with the CS dynamics generated by a CS vector orthogonal to the flat boundary which perfectly guaranties the gauge invariance while provides the Lorentz symmetry violation in the Maxwell-Chern-Simons part coherent with boundary implementation. On the other half-space we have a massive quantum vector
field in vacuum.
Sect. 2 is devoted to definitions: namely, real massive Abelian vector fields are put within a
pseudo-scalar background with a constant gradient on an half-space, while in the complementary
half-space they propagate in an empty Minkowski space. Then the quantum theory for such a kind of
configuration for the vector fields is briefly outlined. In Sect. 3 the Bogolyubov transformations
connecting the operator algebras of creation and destruction operators are derived (in a full analogy with their usage for quantization of matter in curved spaces or accelerated frames, see \cite{birrell}),
which are related to the presence of a boundary  between a parity breaking Chern-Simons background medium and an empty Minkowski space.
In Sect. 4 the functional squeeze operator algebra is obtained and discussed, which allows for
a description and implementation of the transmission and reflection through a boundary
in terms of a purely algebraic approach. A short description of this algebra was presented in \cite{AKStmph}. In Sect. 5  the rate of squeezed pairs
emission and absorption in the presence of a parity breaking Chern-Simons background
medium is evaluated, i.e. the calculation of the probability to find any number of squeezed pairs of vector
particles of the Proca-St\"uckelberg or Chern-Simons kind in crossing the boundary between
the empty space-time and the parity breaking medium is performed. It is an extension of the
Sauter-Schwinger-Nikishov formula \cite{Soldati2011}--\cite{SS}. The Appendix contains some further
technical details concerning the above mentioned computation of the emission-absorption rate.
Throughout this paper we shall use a Minkowski metric tensor
$ g_{\mu\nu}=g^{\,\mu\nu} =\mathrm{diag}(+1,-1,-1,-1)$ and
a natural system of units $ \hslash=c=1 $, unless explicitly stated.
\section{Massive Vector Fields in a Pseudo-Scalar Background}
We start from the St\"uckelberg-type Lagrangian \cite{stuec,BF} which describes the propagation of an Abelian massive real vector field in the presence of a  pseudo-scalar, or axion-like, background field \cite{Carroll:1989vb}, viz.,
\begin{eqnarray}
{\mathcal L} &=&  -\,{\textstyle\frac14}\,F^{\alpha\beta}(x)F_{\alpha\beta}(x)
-\,{\textstyle\frac14}\,g\,\eta_{c\ell}(x)\,F^{\mu\nu}(x)\widetilde F_{\mu\nu}(x)\,M^{-1}
\nonumber\\
&+&  {\textstyle\frac12}\,m^2\,A_\nu(x)A^\nu(x) + A^\mu(x)\,\partial_\mu B(x) +
{\textstyle\frac12}\,\varkappa\,B^2(x)
\label{lagrangian1}
\end{eqnarray}
where $A^{\mu}$ and $\eta_{c\ell}$ stand for the vector and
background  pseudo-scalar fields respectively,
$\widetilde F^{\mu\nu}={\textstyle\frac12}\,
\varepsilon^{\,\mu\nu\rho\sigma}\,F_{\,\rho\sigma}$ is the dual field strength,
while $B$ is the auxiliary St\"uckelberg scalar field \cite{stuec} with $\varkappa\in\mathbb R$. The positive dimensionless coupling
$g>0$ and the (large) mass parameter $M\gg m$ do specify the intensity and the scale of the pseudo-scalar-vector interaction. For example, axions or axion-like particles arise generically
from string-like models or Standard Model extensions,
with a natural size of the decay constant $ M $,
often also denoted by $ f_{a} $, typically varying between 10$ ^{9} $ and 10$ ^{17} $ GeV
\cite{axionreview}, a pretty large scale that will be suitably used and recognized in the sequel
as an effective UV cut-off
regulator which characterizes the adiabatic approximation of a constant LIV background
vector $ \zeta^{\mu}$ as we shall see hereafter.
Notice that we have included
the Proca mass term for the vector field because, as it is
discussed in \cite{aaep}, the latter is required to account for the strong interaction effects in heavy ions collisions supported by massive vector mesons ($\rho,\omega,\ldots$) in addition to photons.
Moreover, as thoroughly debated in \cite{AACGS2010,mauro}, the mass term for the vector field
appears to be generally necessary to render the dynamics self-consistent in the presence of a
Chern-Simons Lagrangian and is actually induced by the radiative corrections from the
LIV fermionic matter.
The auxiliary part of the St\"uckelberg Lagrangian,
which further violates gauge invariance beyond the mass term for the vector field,
has been introduced to provide -- just owing to the renowned St\"uckelberg trick --
the simultaneous occurrences of power counting renormalizability and perturbative unitarity
for a general interacting theory. Its presence allows for a smooth mass-less limit of the
quantum vector field.
Here we shall consider the adiabatic limit of  a slowly varying classical pseudo-scalar
 background of the kind
\begin{equation}
\eta_{c\ell}(x)\simeq\frac{M}{g}\,\zeta_\lambda x^{\lambda}\,\theta(-\,\zeta\cdot x)
\label{background_eta}
\end{equation}
where $\theta(\cdot)$ is the Heaviside step distribution,
in which a fixed constant four vector $\zeta^\mu$ with the dimensions of an inverse length has been introduced, in a way to violate Lorentz and CPT invariances in the Minkowski half-space
$\zeta\cdot x<0\,.$ In what follows we shall suppose that $\zeta^2\not=0\,.$
If we now insert the specific form (\ref{background_eta}) of the pseudo-scalar background
in the pseudo-scalar-vector coupling Lagrangian we can write
\begin{eqnarray}
-\,{\textstyle\frac14}\,F^{\mu\nu}(x)\widetilde F_{\mu\nu}(x)\,\zeta_\lambda x^{\lambda}\,\theta(-\,\zeta\cdot x)
&=& {\textstyle\frac12}\,\zeta_\mu A_\nu(x)\widetilde F^{\mu\nu}(x)\,\theta(-\,\zeta\cdot x)\nonumber\\
&-& \partial_\mu \left[\,{\textstyle\frac12}\,A_\nu(x)\widetilde F^{\mu\nu}(x)\,
\zeta_\lambda x^{\lambda}\,\theta(-\,\zeta\cdot x)\,\right]
\end{eqnarray}
The very last term in the RHS of the above equality is evidently a boundary term, its contribution to the Action
being reduced for the Gau\ss\ theorem to
\begin{displaymath}
\int_\Omega\mathrm{d}^4x\;\partial_\mu \left[\,{\textstyle\frac12}\,A_\nu(x)\widetilde F^{\mu\nu}(x)\,
\zeta_\lambda x^{\lambda}\,\theta(-\,\zeta\cdot x)\,\right]
={\textstyle\frac12}\int_{\partial\Omega}\mathrm{d}\sigma_\mu\;A_\nu(x)\widetilde F^{\mu\nu}(x)\,
\zeta_\lambda x^{\lambda}\,\theta(-\,\zeta\cdot x)
\end{displaymath}
where $\Omega$ is an arbitrary domain of the Minkowski space that is bounded by the \textit{initial}
and \textit{final} three dimensional
space-like oriented surfaces $\partial\Omega=\Sigma_\imath\cup\Sigma_{f}\,.$
Hence the boundary term won't contribute to the Euler-Lagrange field equations iff
\[\left.A_\nu(x)\widetilde F^{\mu\nu}(x)\,
\zeta_\lambda x^{\lambda}\,\theta(-\,\zeta\cdot x)\right|_{\Sigma_\imath}
=\left.A_\nu(x)\widetilde F^{\mu\nu}(x)\,
\zeta_\lambda x^{\lambda}\,\theta(-\,\zeta\cdot x)\right|_{\Sigma_f}\equiv0\]
which entails a particular fall down of the vector potential and field strength
for large space-like separations in the half-space $\zeta\cdot x<0\,.$
In such a circumstance we can derive the field equations from the equivalent Lagrangian
\begin{eqnarray}
{\mathcal L} &=&  -\,{\textstyle\frac14}\,F^{\alpha\beta}(x)F_{\alpha\beta}(x)
+ {\textstyle\frac12}\,\zeta_\mu A_\nu(x)\widetilde F^{\mu\nu}(x)\,\theta(-\,\zeta\cdot x)
\nonumber\\
&+&  {\textstyle\frac12}\,m^2\,A_\nu(x)A^\nu(x) + A^\mu(x)\,\partial_\mu B(x) +
{\textstyle\frac12}\,\varkappa\,B^2(x)\label{lagrangian2}
\end{eqnarray}
in which the gauge invariance is explicitly broken by all the terms but the first one,
 i.e. the Maxwell's Lagrangian for the radiation fields. Then the Euler-Lagrange field equations read
\begin{eqnarray}
\left\lbrace
\begin{array}{cc}
\partial_\lambda F^{\,\lambda\nu}
+\ m^2\,A^\nu + \zeta_\alpha\widetilde F^{\,\alpha\nu} +
\partial^{\,\nu} B=0 & \qquad{\rm for}\ \zeta\cdot x<0\\
\partial_\lambda F^{\,\lambda\nu}
+\ m^2\,A^\nu + \partial^{\,\nu} B=0 & \qquad{\rm for}\ \zeta\cdot x>0\\
\partial_\nu A^\nu\;=\;\varkappa\,B
\end{array}\right. \label{Euler_Lagrange}
\end{eqnarray}
After contraction of the first pair of the above set of field equations with $\partial_\nu$ we find
\begin{equation}
\left(\Box+\varkappa\,m^{2}\right)B(x)=0
\label{aux}
\end{equation}
whence it follows that the auxiliary St\"uckelberg field is always a decoupled
non-physical real scalar field, which is never affected by the pseudo-scalar
classical background $\forall\,\varkappa\in\mathbb R$. From now on we shall select
the simplest choice $\varkappa=1$ that leads to the Klein-Gordon equation
for the auxiliary field, together with
\begin{eqnarray}
\left\lbrace
\begin{array}{cc}
\Box A^\nu(x)  +  m^2\,A^\nu(x) = \varepsilon^{\,\nu\alpha\rho\sigma}\,\zeta_\alpha\,\partial_\rho A_\sigma(x)
& \qquad{\rm for}\ \zeta\cdot x<0\\
\Box A^\nu(x)  +  m^2\,A^\nu(x)=0
& \qquad{\rm for}\ \zeta\cdot x>0\\
\partial_\nu A^\nu(x)=B(x)
&\qquad\quad \left(\Box+m^{2}\right)B(x)=0
\end{array}\right. \label{EL2}
\end{eqnarray}

In order to find the general solution of the above linear equations (\ref{Euler_Lagrange})
we turn to the momentum space
\[
A^\nu(x)=\int\frac{{\rm d}^4k}{(2\pi)^{3/2}}\,{\rm \mathrm{a}}^\nu(k)\,{\rm e}^{\,-i k\cdot x}
\qquad\qquad
B(x)=\int\frac{{\rm d}^4k}{(2\pi)^{3/2}}\,{\rm b}(k)\,{\rm e}^{\,-ik\cdot x}
\]
so that
\begin{equation}
\left\lbrace
\begin{array}{cc}
\left[\,g^{\,\lambda\nu}\left(\,k^2-m^2\,\right) - k^{\,\lambda}k^{\,\nu} +
i\,\varepsilon^{\,\lambda\nu\alpha\beta}\,\zeta_\alpha\,k_\beta\,\right]
 \mathrm{a}_\lambda(k) + i\,k^{\,\nu}\,\mathrm b(k)=0
& \qquad{\rm for}\ \zeta\cdot x<0\\
\left[\,g^{\,\lambda\nu}\left(\,k^2-m^2\,\right) - k^{\,\lambda}k^{\,\nu}\,\right]
 \mathrm{a}_\lambda(k) + i\,k^{\,\nu}\,\mathrm b(k)=0
& \qquad{\rm for}\ \zeta\cdot x>0\\
k^{\,\lambda}\,\mathrm{a}_\lambda(k) = i\,\mathrm b(k)
\qquad\qquad\quad \left(k^2-m^{2}\right)\mathrm b(k)=0
\end{array}\right.
\label{MEL2}
\end{equation}
The general solution for the auxiliary field operator is well known, \textit{viz.,}
\begin{eqnarray}
B(x)=\int\mathrm{d}\mathbf{k}\,[\,\mathrm{b}_{\bf k}\,u_{\bf k}(x)+
\mathrm{b}^{\dagger}_{\bf k}\,u^{\ast}_{\bf k}(x)\,]\\
u_{\bf k}(x)=[\,(2\pi)^{3}\,2k_{0}\,]^{-1/2}\,\exp\{-\,i\,k_{\nu}x^{\nu}\}
\qquad\quad k_{0}=\sqrt{\mathbf{k}^{2}+m^{2}}
\end{eqnarray}
with ghost-like canonical commutation relations
\begin{equation}
[\,\mathrm{b}^{\dagger}_{\bf k}\,,\,\mathrm{b}_{\bf p}\,]=\delta(\mathbf{k}-\mathbf{p})\qquad\quad
[\,\mathrm{b}_{\bf k}\,,\,\mathrm{b}_{\bf p}\,]=0
\end{equation}
The general solutions on the two half-spaces separated by the hyper-plane
$ \zeta\cdot x=0 $ have to be set in a more
convenient form to the purpose of our discussion. It turns out that
the Proca-St\"{u}ckelberg and the Maxwell-Chern-Simons massive vector fields
face one each other at the boundary $\zeta\cdot  x=0$.
Hence, continuity of the quantum fields does require
equality on the surface separating the parity breaking medium from the Minkowski space: namely,
\begin{equation}
\delta(\zeta\cdot x)\left[ \,A^{\mu}_{\rm PS}(x)-A^{\mu}_{\rm CS}(x)\,\right] = 0
\label{boundary}
\end{equation}
while the auxiliary non-physical field $B(x)$ is by no means affected by the presence of the
hyper-plane $\zeta\cdot x=0\,,$ as already noticed.
\subsection{The Proca-St\"uckelberg Quantum Field in the Presence of a Boundary}
On the half-space $\zeta\cdot x>0$, the general solutions of the field equations are the well known
Proca-St\"uckelberg vector and auxiliary ghost scalar free quantum fields.
 In order to write  the general solutions in the presence of the infinite hyper-plane
 separation boundary $\zeta\cdot x=0$,
from now on it is convenient to use a slightly different notation: namely,
we set $ \mathrm{x}=(x_{0},x^{1},x^{2},x^{3})=(t,x,y,z) $ in natural units.
Consider the case of the spatial Chern-Simons constant vector $\zeta^{\,\mu}=(0,\zeta,0,0)$ with
$ \zeta>0 $ and
the corresponding space-like hyper-plane $ x^1\equiv x=0\,,$ in such a manner that
$\delta(\zeta\cdot\mathrm  x)=\zeta^{\,-1}\delta(x)\,.$
Let us define the following quantities:
\begin{eqnarray*}
\hat{\mathrm k} \equiv (k_0, k_y, k_z);\quad
\hat{\mathrm x} \equiv (t, y, z);\quad
\hat{\mathrm k} \cdot \hat{\mathrm x} = t k_0 - yk_y - zk_z;\quad
k_0=\omega_{\,\mathbf{k}}\equiv\sqrt{\mathbf k^2+m^2}
\end{eqnarray*}
Then we can write the Proca-St\"{u}ckelberg solution in form
\begin{eqnarray}
A^{\mu}_{\rm PS}(x,\hat{\mathrm x}) &=& \int\mathrm d\mathbf k\sum_{r=1}^3\
\left[\,\mathrm{a}_{\,{\bf k},\,r}\,u^{\,\mu}_{\,{\bf k}\,,\,r}(x,\hat{\mathrm x})
+ \mathrm{a}^\dagger_{\,{\bf k},\,r}\,
u^{\,\mu\,\ast}_{\,{\bf k}\,,\,r}(x,\hat{\mathrm x})\,\right]
\\
u^{\,\nu}_{\,{\bf k}\,,\,r}(x,\hat{\mathrm x})&=& \theta(-\,x)\,[\,(2\pi)^3\,2\omega_{\,\mathbf{k}}\,]^{-1/2}\,
e_{\,r}^{\,\nu}({\bf k})\,
\exp\{\,i\,K x - i\,\hat{\mathrm k}\cdot\hat{\mathrm x}\};\quad (\,r=1,2,3\,)\nonumber
\end{eqnarray}
with $x\le 0\,,\, K\equiv k^1=k_x\,, $
where the creation destruction operators fulfill the canonical commutation relations
\begin{equation}
 [\,\mathrm a_{\,{\bf k},\,r}\,,\,\mathrm a^\dagger_{\,{\bf k}^\prime,\,s}\,]=
 \delta({\bf k}-{\bf k}^{\,\prime})\,\delta_{rs}
\qquad\quad
\end{equation}
all the remaining commutators being equal to zero.
The three linear polarization real vectors do satisfy the orthogonality and closure relations
on the mass shell $K^{\,2}=\hat{\mathrm k}^2-m^2\,:$ namely,
\begin{eqnarray*}
\mathrm k_{\mu}\,e_{\,r}^{\,\mu}({\bf k})=0;\qquad
 -\,g_{\,\mu\nu}\;e_{\,r}^{\,\mu}({\bf k})\,
e_{\,s}^{\,\nu}({\bf k})=\delta_{\,rs};
\qquad
\sum_{r=1}^3\,e_{\,r}^{\,\mu}({\bf k})\,
e_{\,r}^{\,\nu}({\bf k})
= -\;g^{\;\mu\nu} + \frac{\mathrm k^{\mu}\,\mathrm k^{\nu}}{m^2}
\end{eqnarray*}
It is physically motivated
to split the Proca-St\"uckelberg vector field into
the so called progressive (or incident) and regressive (or reflected) parts, that corresponds to positive and negative
longitudinal momenta $ K=k^1=k_x \gtrless 0$ respectively:  namely,
\begin{eqnarray*}
A_{\,\rightarrow}^{\mu}(x,\hat{\mathrm x})\equiv
\int_0^\infty\mathrm dK\int_{-\infty}^{\infty}\mathrm dk_y\int_{-\infty}^{\infty}\mathrm dk_z
\sum_{r=1}^3\
\left[\,\mathrm{a}_{\,{\bf k},\,r}\,u^{\,\mu}_{\,{\bf k}\,,\,r}(x,\hat{\mathrm x})
+ \mathrm{a}^\dagger_{\,{\bf k},\,r}\,
u^{\,\mu\,\ast}_{\,{\bf k}\,,\,r}(x,\hat{\mathrm x})\,\right]\\
A_{\,\leftarrow}^{\nu}(x,\hat{\mathrm x})\equiv
\int_0^\infty\mathrm dK\int_{-\infty}^{\infty}\mathrm dk_y\int_{-\infty}^{\infty}\mathrm dk_z
\sum_{r=1}^3\
\left[\,\mathrm{a}_{\,-{\bf k},\,r}\,u^{\,\nu}_{\,-{\bf k}\,,\,r}(x,\hat{\mathrm x})
+ \mathrm{a}^\dagger_{\,-{\bf k},\,r}\,
u^{\,\nu\,\ast}_{\,-{\bf k}\,,\,r}(x,\hat{\mathrm x})\,\right]
\end{eqnarray*}
where we understand $ -\mathbf{k}\equiv(-K,k_y,k_z)\,. $ Of course we have
\begin{eqnarray*}
&&u^{\,\nu}_{\,- {\bf k},\,r}(x,\hat{\mathrm x})=\theta(-x)
[\,(2\pi)^3\,2k_0\,]^{-1/2}\,
e_{\,r}^{\,\nu}(-{\bf k})\,
\exp\{\,-\,iK x + iyk_{y} + izk_{z} - itk_0\}\\
&&\equiv\bar u^{\,\nu}_{\,{\bf k},\,r}(x,\hat{\mathrm x})
\end{eqnarray*}
\begin{equation}
[\,\mathrm a_{\,-{\bf k},\,r}\,,\,\mathrm a^\dagger_{\,{\bf k},\,s}\,]=
[\,\mathrm a_{\,-{\bf k},\,r}\,,\,\mathrm a_{\,{\bf k},\,s}\,]=0
\end{equation}
in such a manner that if we set
\begin{equation}
\mathrm a_{\,- {\bf k},\,r}\equiv\bar{\mathrm a}_{\,{\bf k},\,r}\qquad\quad
\forall\,\mathbf k=(K,k_y,k_z)\quad {\rm with}\quad K>0\,,\,k_y,k_z\in\mathbb R
\end{equation}
we can eventually write the normal modes expansions for $x\le 0$
\begin{eqnarray}
A_{\,\rightarrow}^{\mu}(x,\hat{\mathrm x})\equiv
\int_0^\infty\mathrm dK\int_{-\infty}^{\infty}\mathrm dk_y\int_{-\infty}^{\infty}\mathrm dk_z
\sum_{r=1}^3\
\left[\,\mathrm{a}_{\,{\bf k},r}\,u^{\,\mu}_{{\bf k}, r}(x,\hat{\mathrm x})
+ \mathrm{a}^\dagger_{{\bf k},r}\,
u^{\,\mu\,\ast}_{{\bf k},r}(x,\hat{\mathrm x})\right]
\label{Aprogre} \\
A_{\,\leftarrow}^{\nu}(x,\hat{\mathrm x})\equiv
\int_0^\infty\mathrm dK\int_{-\infty}^{\infty}\mathrm dk_y\int_{-\infty}^{\infty}\mathrm dk_z
\sum_{r=1}^3\
\left[\bar{\mathrm a}_{{\bf k},r}\,\bar u^{\nu}_{{\bf k},r}(x,\hat{\mathrm x})
+ \bar{\mathrm a}^\dagger_{{\bf k},r}\,
\bar u^{\nu\ast}_{{\bf k},r}(x,\hat{\mathrm x})\right]
\label{Aregre}
\end{eqnarray}
where
\begin{equation}
[\,\bar{\mathrm a}_{\,{\bf k},\,r}\,,\,\mathrm a^\dagger_{\,{\bf k}',\,s}\,]=
[\,\bar{\mathrm a}_{\,{\bf k},\,r}\,,\,\mathrm a_{\,{\bf k}',\,s}\,]=0
\end{equation}
so that
\begin{equation}
[\,A_{\,\rightarrow}^{\mu}(x,\hat{\mathrm x})\,,\,
A_{\,\leftarrow}^{\nu}(x^{\,\prime},\hat{\mathrm x}^{\,\prime})\,]=0
\end{equation}
Now it is expedient to change the integration variable into the normal modes expansions
(\ref{Aprogre},\ref{Aregre}) from $ K>0 $ to the positive frequency
$ \omega(K)=\sqrt{K^{\,2}+k_y^2+k_z^2+m^2}\ge m\,.$ After setting $ \hat{\mathrm{k}}=(\omega,k_y,k_z) $
we get
\begin{eqnarray*}
&&A_{\,\rightarrow}^{\mu}(x,\hat{\mathrm x}) \equiv
\int_m^\infty\omega\,\mathrm d\omega
\int\!\!\!\!\int_{-\infty}^{\infty}\mathrm dk_y\,\mathrm dk_z\;
\frac{\theta(\,\hat{\mathrm{k}}^2 -m^2\,)}{\sqrt{\hat{\mathrm{k}}^2 -m^2}}
\sum_{r=1}^3\mathrm{a}_{\,{\bf k},\,r}\,u^{\,\mu}_{\,{\bf k}\,,\,r}(x,\hat{\mathrm x})
+\ {\rm H.c.}\\
&&=\int_m^\infty\mathrm d\omega\int\!\!\!\!\int_{-\infty}^{\infty}\mathrm dk_y\,\mathrm dk_z\,
\theta(\,\hat{\mathrm{k}}^2 -m^2\,)\sum_{r=1}^3\left[ \,
\mathrm a_{\,\hat{\rm k},\,r}\,u^{\,\mu}_{\,\hat{\rm k}\,,\,r}(x,\hat{\mathrm x})
+ \mathrm a^{\,\dagger}_{\,\hat{\rm k},\,r}\,u^{\,\mu\,\ast}_{\,\hat{\rm k}\,,\,r}(x,\hat{\mathrm x})\,\right]
\end{eqnarray*}
where
\begin{equation}
u^{\,\mu}_{\,{\hat{\rm k}}\,,\,r}(x,\hat{\mathrm x})=\frac{\theta(-x)\,e_{\,r}^{\,\mu}(\hat{\rm k})}
{\surd\,[\,(2\pi)^3\,2K(\hat{\mathrm{k}})\,]}\;
\exp\left\lbrace -\,i\,\omega\,t + iyk_y + izk_z + i\,x\,\sqrt{\hat{\mathrm{k}}^{2}- m^2}\,\right\rbrace
\label{PSprogressive}
\end{equation}
with
\begin{equation}
e_{\,r}^{\,\mu}(\hat{\rm k})=e_{\,r}^{\,\mu}(K(\hat{\mathrm{k}}),k_y,k_z)\qquad\quad
(\,r=1,2,3\,)\qquad\quad \mathrm{a}_{\,{\bf k},\,r}\;\sqrt{\frac{\omega}{K}}={\mathrm a}_{\,{\hat{\rm k}},\,r}
\end{equation}
in such a manner that we come to the canonical commutation relations
\[
[\,{\mathrm a}_{\,{\hat{\rm k}},\,r}\,,\,{\mathrm a}_{\,{\hat{\rm k}^{\,\prime}},\,s}\,]=0
\qquad\quad
[\,{\mathrm a}^{\,\dagger}_{\,{\hat{\rm k}},\,r}\,,\,{\mathrm a}^{\,\dagger}_{\,{\hat{\rm k}^{\,\prime}},\,s}\,]=0
\]
\[
[\,{\mathrm a}_{\,{\hat{\rm k}},\,r}\,,\,{\mathrm a}^{\,\dagger}_{\,{\hat{\rm k}^{\,\prime}},\,s}\,]
=\frac{\omega}{K}\;\delta(K-K^{\,\prime})\,\delta(k_y-k_y^{\,\prime})\,\delta(k_z-k_z^{\,\prime})
=\delta(\omega-\omega^{\,\prime})\,\delta(k_y-k_y^{\,\prime})\,\delta(k_z-k_z^{\,\prime})
\]
with $\omega,\omega^{\,\prime}\ge0\ $.
Notice that the incident or progressive wave functions $ u^{\,\mu}_{\,{\hat{\rm k}}\,,\,r}(x,\hat{\mathrm x}) $
are tempered distributions satisfying the non-homogeneous Klein-Gordon equation
\begin{eqnarray*}
(\,\square+m^2\,)\,u^{\,\mu}_{\,{\hat{\rm k}}\,,\,r}(x,\hat{\mathrm x})=
-\left\lbrace\delta^{\,\prime}(x)+i\,\delta(x)\,\sqrt{\hat{\mathrm{k}}^{2}- m^2}\,\right\rbrace
\frac{e_{\,r}^{\,\mu}(\hat{\rm k})}
{\surd\,[\,(2\pi)^3\,2K(\hat{\mathrm{k}})\,]}\;
\exp\lbrace -\,i\,\hat{\mathrm{k}}\cdot \hat{\mathrm{x}}\rbrace
\end{eqnarray*}
By repeating the very same manipulations to the regressive part $ A_{\,\leftarrow}^{\mu}(x,\hat{\mathrm x}) $
of the Proca-St\"uckelberg vector field, we eventually come to the normal modes expansion for $ x\le 0 $
\begin{eqnarray*}
A^{\nu}_{\,\leftarrow}(\mathrm x)=\int_m^\infty\mathrm d\omega\int\!\!\!\!\int_{-\infty}^{\infty}\mathrm dk_y\,\mathrm dk_z\,\theta(\,\hat{\mathrm{k}}^2 -m^2\,)\sum_{r=1}^3\,
\left[\,\bar{\mathrm{a}}_{\,\hat{\rm k},\,r}\,\bar u_{\,\hat{\rm k},\,r}^{\,\nu}(\mathrm x) +
\bar{\mathrm{a}}_{\,\hat{\rm k},\,r}^{\,\dagger}\,\bar u_{\,\hat{\rm k},\,r}^{\,\nu\,\ast}(\mathrm x)\,\right]
\end{eqnarray*}
where the reflected or regressive wave functions have the form
\begin{eqnarray}
\bar{u}^{\,\mu}_{\,{\hat{\rm k}},\,r}({\mathrm x})=\theta(-x)\,
[\,(2\pi)^3\,2K(\hat{\mathrm{k}})\,]^{-1/2}\,\bar{e}_{\,r}^{\,\mu}(\,\hat{\rm k})\,
\exp\left\lbrace -\,i\,x\,K(\hat{\mathrm{k}}) - i\,\hat{\rm k}\cdot\hat{\rm x}\right\rbrace\\
\bar{e}_{\,r}^{\,\mu}(\hat{\rm k})=e_{\,r}^{\,\mu}(-\,K(\hat{\mathrm{k}}),k_y,k_z)\qquad\quad
(\,r=1,2,3\,)
\end{eqnarray}
and do fulfill the non-homogeneous Klein-Gordon equation
\begin{eqnarray*}
(\,\square+m^2\,)\,\bar u^{\,\mu}_{\,{\hat{\rm k}}\,,\,r}(x,\hat{\mathrm x})=
-\left\lbrace\delta^{\,\prime}(x)-i\,\delta(x)\,\sqrt{\hat{\mathrm{k}}^{2}- m^2}\,\right\rbrace
\frac{\bar e_{\,r}^{\,\mu}(\hat{\rm k})}
{\surd\,[\,(2\pi)^3\,2K(\hat{\mathrm{k}})\,]}\;
\exp\lbrace -\,i\,\hat{\mathrm{k}}\cdot \hat{\mathrm{x}}\rbrace
\end{eqnarray*}
while the corresponding creation and annihilation operators satisfy canonical commutation relations
\begin{equation}
[\,\bar{{\mathrm a}}_{\,{\hat{\rm k}},\,r}\,,\,{\mathrm a}_{\,{\hat{\rm k}^{\,\prime}},\,s}\,]=
[\,{\mathrm a}^{\,\dagger}_{\,{\hat{\rm k}},\,r}\,,\,\bar{{\mathrm a}}_{\,{\hat{\rm k}^{\,\prime}},\,s}\,]=0
\qquad\quad
[\,\bar{{\mathrm a}}_{\,{\hat{\rm k}},\,r}\,,\bar{{\mathrm a}}^{\,\dagger}_{\,{\hat{\rm k}^{\,\prime}},\,s}\,]=
\delta(\,\hat{\rm k}-\hat{\rm k}^{\,\prime}\,)\,\delta_{\,rs}
\end{equation}
It's worth noticing that all the wave functions are normalized in order to reproduce the
constant vector current: for example, for incident or progressive wave functions we find
\begin{equation}
(2\pi)^3\, g_{\mu\nu}\,
u^{\,\mu\,\ast}_{\,{\hat{\rm k}},\,s}(\mathrm x)\,
i\!\buildrel \!\leftrightarrow\over {\partial_\lambda}\!
 u^{\,\nu}_{\,{\hat{\rm k}},\,r}(\mathrm x)=\delta_{rs}\,\frac{k_{\lambda}}{K}
 \qquad\quad k^{\,\mu}=\left(\omega,K=\sqrt{\hat{\rm k}^2 - m^2},k_y,k_z\right)
\end{equation}
The normalization of the incident or progressive vector plane waves is such that
\begin{eqnarray}
g_{\mu\nu}\int\mathrm d\hat{\rm x}\;\left.
u^{\,\mu\,\ast}_{\,\hat{\rm p},r}(x,\hat{\rm x})\,i\!\buildrel \!\leftrightarrow\over {\partial_\lambda}
u_{\,\hat{\rm k},s}^{\,\nu}(x,\hat{\rm x})\,\right\rfloor_{\,x\,=\,0}
=\delta(\hat{\rm k}-\hat{\rm p})\,\delta_{rs}\;\frac{k_{\lambda}}{K}\\
g_{\mu\nu}\int\mathrm d\hat{\rm x}\;\left.
u^{\,\mu}_{\,\hat{\rm p},r}(x,\hat{\rm x})\,i\!\buildrel \!\leftrightarrow\over {\partial_\lambda}
u_{\,\hat{\rm k},s}^{\,\nu\,\ast}(x,\hat{\rm x})\,\right\rfloor_{\,x\,=\,0}
= \delta(\hat{\rm k}-\hat{\rm p})\,\delta_{rs}\;\left( -\,\frac{k_{\lambda}}{K}\right) \\
g_{\mu\nu}\int\mathrm d\hat{\rm x}\;\left.
u^{\,\mu}_{\,\hat{\rm p},r}(x,\hat{\rm x})\,i\!\buildrel \!\leftrightarrow\over {\partial_\lambda}
u_{\,\hat{\rm k},s}^{\,\nu}(x,\hat{\rm x})\,\right\rfloor_{\,x\,=\,0}=0
\end{eqnarray}
in which $ \upsilon=\omega/K $ is the (phase) velocity of the vector plane wave
$u^{\,\nu}_{\,{\hat{\rm k}},\,r}(\mathrm{x})\,.$ Now, concerning the reflected or regressive plane waves,
it is convenient to set
\begin{eqnarray}
{u}^{\,\mu}_{-\,{\hat{\rm k}},\,r}({\mathrm x})=\theta(-x)\,
[\,(2\pi)^3\,2K(\hat{\mathrm{k}})\,]^{-1/2}\,{e}_{\,r}^{\,\mu}(-\,\hat{\rm k})\,
\exp\left\lbrace -\,i\,x\,K(\hat{\mathrm{k}}) + i\,\hat{\rm k}\cdot\hat{\rm x}\right\rbrace\\
{e}_{\,r}^{\,\mu}(-\,\hat{\rm k})=e_{\,r}^{\,\mu}(-\,K(\hat{\mathrm{k}}),-k_y,-k_z)\qquad\quad
(\,r=1,2,3\,)
\end{eqnarray}
in such a manner that we can recast the normal modes expansion of the Proca-St\"uckelberg
vector field in the simple form
\begin{eqnarray}
A^{\mu}_{\rm PS}(x,\hat{\mathrm{x}})=A^{\mu}_{\rightarrow}(x,\hat{\mathrm{x}})
+A^{\mu}_{\leftarrow}(x,\hat{\mathrm{x}})=
\int\mathrm{d}\hat{\mathrm{k}}\sum_{r=1}^3\mathrm{a}_{\,\hat{\mathrm{k}},\,r}\,
u^{\,\mu}_{\,\hat{\mathrm{k}}\,,\,r}(x,\hat{\mathrm x})+\ {\rm H.c.}\\
\int\mathrm{d}\hat{\mathrm{k}}\equiv
\int_{-\infty}^\infty\mathrm d\omega\,\theta(\omega^2-m^2)
\int\!\!\!\!\int_{-\infty}^{\infty}\mathrm dk_y\,\mathrm dk_z\,\theta(\,\hat{\mathrm{k}}^2 -m^2\,)
\end{eqnarray}
with the canonical commutation relations
\[
[\,{\mathrm a}_{\,{\hat{\rm k}},\,r}\,,\,{\mathrm a}_{\,{\hat{\rm k}^{\,\prime}},\,s}\,]=0
\qquad\quad
[\,{\mathrm a}^{\,\dagger}_{\,{\hat{\rm k}},\,r}\,,\,{\mathrm a}^{\,\dagger}_{\,{\hat{\rm k}^{\,\prime}},\,s}\,]=0
\qquad\quad {\hat{\rm k}}\in\mathbb{R}^{3}
\]
\[
[\,{\mathrm a}_{\,{\hat{\rm k}},\,r}\,,\,{\mathrm a}^{\,\dagger}_{\,{\hat{\rm k}^{\,\prime}},\,s}\,]
=\delta(\omega-\omega^{\,\prime})\,\delta(k_y-k_y^{\,\prime})\,\delta(k_z-k_z^{\,\prime})
\qquad\quad\omega^{\,2}\ge m^{2}\,\vee\,k_y,k_z\in\mathbb{R}
\]
On the other hand, the general solutions for $\zeta\cdot x<0$, concerning the Maxwell-Chern-Simons
free quantum field, have been extensively discussed and applied in
\cite{AACGS2010,axion1,axion2} for the massive case and in \cite{AGS2002} for the massless case.
However, in the light of the present applications, it is better to shortly overview this important topic.
\subsection{Maxwell-Chern-Simons Quantum Field}
In the following we shall review and extend the main tools developed in
\cite{AKS11}, which are necessary to understand the squeezed pairs emission and
absorption in the presence of a parity breaking background medium.
Let us first recall the construction of the so called
{\sf chiral or birefringent polarization vectors}
for the Maxwell-Chern-Simons (MCS) vector field. Here we aim to develop a rather general frame which
could allow for an understanding of  the kinematics of massive vector particles in the presence of a space-like Chern-Simons LIV vector. The starting point is the rank-two symmetric matrix
\cite{AACGS2010}
\begin{equation}
S^{\nu}_{\phantom{\nu}\lambda}\equiv\varepsilon^{\mu\nu\alpha\beta}\,\zeta_{\alpha}\,k_{\beta}\,
\varepsilon_{\mu\lambda\rho\sigma}\,\zeta^{\,\rho}k^{\sigma}=
\delta^{\,\nu}_{\;\lambda}\,{\mbox{\tt D}}
+ k^{\nu}\,k_{\lambda}\,\zeta^2 + \zeta^{\,\nu}\,\zeta_{\lambda}\,k^2
- \zeta\cdot k\,(\zeta_{\lambda}\,k^{\nu} + \zeta^{\nu}\,k_{\,\lambda})
\end{equation}
where
\[
{\mbox{\tt D}}\;\equiv\;(\zeta\cdot k)^2-\zeta^2\,k^2\;=\;\textstyle\frac12\;S^{\nu}_{\phantom{\nu}\nu} .
\]
It is convenient to introduce the
two orthogonal, one-dimensional, Hermitian projectors
\begin{equation}
\bfpi^{\,\mu\nu}_{\,\pm}\equiv
\frac{S^{\,\mu\nu}}{2\,{\mbox{\tt D}}}\;
\pm\;\frac{i}{2}\,
\varepsilon^{\mu\nu\alpha\beta}\,\zeta_{\alpha}\,k_{\beta}\,{\mbox{\tt D}}^{\,-\frac12}
=\left(\bfpi^{\,\nu\mu}_{\,\pm}\right)^\ast=\left(\bfpi^{\,\mu\nu}_{\,\mp}\right)^\ast
\qquad\quad(\mbox{\tt D}>0)
\end{equation}
A couple of chiral polarization vectors for the Maxwell-Chern-Simons free vector field can be constructed
out of some tetrad of constant quantities $\epsilon_\nu$, taking into account that we have
\begin{eqnarray}
 \bfpi^{\,\mu\lambda}_{\,\pm}\,\epsilon_\mu\epsilon_\lambda=
{\mbox{\tt D}}\,\epsilon^2+\zeta^2(\epsilon\cdot k)^2=
[\,(\zeta\cdot k)^2-\zeta^2\,k^2\,]\,\epsilon^2+\zeta^2(\epsilon\cdot k)^2
\end{eqnarray}
For example, for the spatial Chern-Simons vector
$\zeta^{\mu}=(0,\zeta_{\,x},0,0)$
 we can always build up a pair of space-like, complex, chiral polarization vectors
\begin{equation}
\varepsilon_{\,\pm}^{\,\mu}(k)=
\bfpi^{\,\mu\lambda}_{\,\pm}\,\epsilon_\lambda\,\left[(k_y^2 - k_0^2)/(k_z^2 + k_y^2 - k_0^2)\right]^{-\frac12}.
\label{edef}
\end{equation}
By the very construction, for $\mbox{\tt D}>0$ this couple of chiral polarization vectors satisfy the conjugation and orthogonality relations
\begin{displaymath}
\varepsilon_{\,\pm}^{\,\mu\,\ast}(k)=\varepsilon_{\,\mp}^{\,\mu}(k);\quad
-\,{\textstyle\frac12}\,g_{\mu\nu}\;\varepsilon^{\,\mu\ast}_{\pm}(k)\,\varepsilon^{\,\nu}_{\pm}(k)+{\rm c.\,c.}
=1;\quad
{\textstyle\frac12}\,g_{\mu\nu}\,\varepsilon^{\,\mu\ast}_{\pm}(k)\,\varepsilon^{\,\nu}_{\mp}(k)+{\rm c.\,c.}=0
\end{displaymath}
as well as the closure relations
\begin{equation}
\varepsilon^{\,\mu\ast}_{+}(k)\,\varepsilon^{\,\nu}_{\,+}(k) +
\varepsilon^{\,\mu\ast}_{\,-}(k)\,\varepsilon^{\,\nu}_{-}(k) =
\varepsilon^{\,\mu}_{-}(k)\,\varepsilon^{\,\nu}_{\,+}(k) +
\varepsilon^{\,\mu}_{\,+}(k)\,\varepsilon^{\,\nu}_{-}(k)= {\mbox{\tt D}}^{-1}\,{S^{\,\mu\nu}}
\end{equation}
In order to obtain the normal modes expansion of the MCS
quantum field, let's introduce the kinetic $4\times4$
Hermitian matrix $\mathbb K$ with complex entries
\begin{equation}
K_{\,\lambda\nu}\equiv
g_{\,\lambda\nu}\left(k^2-m^2\right) +
i\varepsilon_{\lambda\nu\alpha\beta}\,\zeta^{\,\alpha} k^{\,\beta}
\label{kdef}
\end{equation}
which satisfy
\[
K_{\,\lambda\nu}=K^{\,\ast}_{\,\nu\lambda}
\]
Now we are ready to find the general solution of the free field equations
(\ref{MEL2}) for $\zeta\cdot x<0$.
From the relationships
(\ref{kdef}) and (\ref{edef}) we readily obtain
\begin{eqnarray}
K^{\,\mu}_{\phantom{\mu}\nu}\,\varepsilon^{\,\nu}_{\pm}(k) &=&
\left[\,\delta^{\,\mu}_{\phantom{\mu}\nu}\left(k^2-m^{2}\right)
+ {\sqrt{\mbox{\tt D}}}\,\left(\,\bfpi^{\,\mu}_{\,+\,\nu}\;-\;\bfpi^{\,\mu}_{\,-\,\nu}\,\right)\,\right]
\varepsilon^{\,\mu}_{\pm}(k)\nonumber\\
&=& \left(k^2-m^{2} \pm\,\sqrt{\mbox{\tt D}} \,\right)\,
\varepsilon^{\,\mu}_{\pm}(k)
\end{eqnarray}
which shows that the polarization vectors of positive and negative chirality are solutions of the vector field equations for $\zeta\cdot x<0$ iff
\begin{eqnarray}
k^{\,\mu}_{\,\pm}=(\omega_{\,{\bf k}\,,\,\pm}\,,\,{\bf k})\qquad\quad
\varepsilon^{\,\mu}_{\pm}({\bf k})=\varepsilon^{\,\mu}_{\pm}(k_{\pm})\qquad\quad
\left(\,k^{0}_{\pm}\;=\;\omega_{\,{\bf k}\,,\,\pm}\,\right)\\
\omega_{\,{\bf k}\,,\,\pm}=
\sqrt{{\bf k}^2+m^{2}+\frac12\zeta_{\,x}^{\,2}\pm\zeta_{\,x}\sqrt{k_x^2+m^2+\frac14\zeta_{\,x}^{\,2}}}
\qquad {\rm for}\quad\zeta^\mu=(0,\zeta_{\,x},0,0).
\label{MCSDR}
\end{eqnarray}
To complete our construction of a basis
we introduce the further pair of orthogonal and suitably normalized polarization vectors,
respectively the so called scalar and longitudinal polarization real vectors
\begin{eqnarray}
\varepsilon^{\,\mu}_{\,S}(k)\;\equiv\;
\frac{k^{\,\mu}}{\sqrt{\,k^2}}\qquad\qquad
(\,k^2>0\,)\\
\varepsilon^{\,\mu}_{\,L}(k)\;\equiv\;\left(\mbox{\tt D}\,k^2\right)^{-\frac12}\left(
k^2\,\zeta^{\,\mu} - k^{\,\mu}\,\zeta\cdot k\,\right)
\qquad\qquad
(\,k^2>0\vee\mbox{\tt D}>0\,)
\end{eqnarray}
which fulfill by construction
\begin{eqnarray}
k_\mu\,\varepsilon^{\,\mu}_{\,L}(k)=0\qquad\quad
k_\mu\,\varepsilon^{\,\mu}_{\,S}(k)=\sqrt{k^2}\qquad(\,k^2>0\,)\\
g_{\mu\nu}\;\varepsilon^{\,\mu}_{\,S}(k)\,\varepsilon^{\,\nu}_{\,S}(k)\ =\ 1
\qquad\quad
g_{\mu\nu}\;\varepsilon^{\,\mu}_{\,L}(k)\,\varepsilon^{\,\nu}_{\,L}(k)\ =\ -\,1\\
g_{\,\mu\nu}\,\varepsilon^{\,\mu}_{\,S}(k)\,\varepsilon^{\,\nu}_{\,L}(k)=
g_{\,\mu\nu}\,\varepsilon^{\,\mu}_{\,S}(k)\,\varepsilon^{\,\nu}_{\pm}(k)=
g_{\,\mu\nu}\,\varepsilon^{\,\mu}_{\,L}(k)\,\varepsilon^{\,\nu}_{\pm}(k)=0
\end{eqnarray}
Hence we have at our disposal $\forall\,k^{\,\mu}$ with $k^2>0 \vee \mbox{\tt D}>0$ a complete
and orthogonal chiral set of four polarization vectors: namely,
\begin{eqnarray}
\varepsilon^{\,\mu}_{\,A}(k)=\left\lbrace
\begin{array}{cc}
{k^{\,\mu}}/{\sqrt{\,k^2}} & {\rm for}\ A=S\\
\left(k^2\,\zeta^{\,\mu} - k^{\,\mu}\,\zeta\cdot k\,\right)/
\displaystyle\sqrt{\mbox{\tt D}\,k^2} & {\rm for}\ A=L\\
\varepsilon^{\,\mu}_{\pm}(k_{\pm}) & {\rm for}\ A=\pm
\end{array}
\right.
\qquad\quad
-(\,k^2>0\,\vee\,\mbox{\tt D}>0\,).
\end{eqnarray}
In order to fully implement the canonical quantum theory of the MCS
massive vector field for the simple choice $\varkappa=1$, it is convenient to introduce
the polarized plane waves according to
\begin{eqnarray}
v_{\,{\bf k}\,A}^{\,\nu}(x) =
\left[\,(2\pi)^3\,2\omega_{\,{\bf k}\,A}\,\right]^{-\frac12}\,\varepsilon^{\,\nu}_{\,A}(k)\
\exp\{-\,i\omega_{\,{\bf k}\,A}\,t + i\,{\bf k}\cdot{\bf x}\}
\end{eqnarray}
where the dispersion relation for the scalar and longitudinal frequencies is
the covariant one, viz.,
\[\omega_{\,{\bf k}\,S}=\omega_{\,{\bf k}\,L}=
\displaystyle\sqrt{{\bf k}^{2}+m^2}\equiv\omega_{\,{\bf k}}\]
so that we can write
\begin{equation}
 k_\nu\,\varepsilon^{\,\nu}_{\,S}(k)=m\qquad\quad
i\partial_\nu v_{\,{\bf k}\,S}^{\,\nu}(x)=u_{\,{\bf k}}(x)
\end{equation}
It follows therefrom that the general solution of the Euler-Lagrange field equations (\ref{EL2})
for the quantum massive vector field when $\varkappa=1$ and $\zeta\cdot x<0$ takes the form
\begin{eqnarray}
A^{\nu}(x) &=& A^{\nu}_{\rm CS}(x)-\partial^{\,\nu}B(x)/m^2\\
A^{\nu}_{\rm CS}(x)&=&
\int\mathrm d\mathbf k\sum_{A=\pm,L}\,
\left[\,c_{\,{\bf k},A}\,v_{\,{\bf k}\,A}^{\,\nu}(x) +
c_{\,{\bf k},A}^{\,\dagger}\,v_{\,{\bf k}\,A}^{\,\nu\,\ast}(x)\,\right]\\
B(x)&=& {m}\int\mathrm d\mathbf k\,
\left[\,b_{\,{\bf k}}\,u_{\,{\bf k}}(x) +
b_{\,{\bf k}}^{\,\dagger}\,u_{\,{\bf k}}^{\ast}(x)\,\right]
\end{eqnarray}
where the canonical commutation relations holds true, viz.,
\begin{eqnarray}
\left[\,c_{\,{\bf k},A}\,,\,c_{\,{\bf k}^{\prime},A^{\prime}}^{\,\dagger}\,\right]\;=\;-\,g_{AA^{\prime}}\,
\delta({\bf k}-{\bf k}^{\prime})\qquad\quad
c_{\,{\bf k},S}=b_{\,{\bf k}}
\end{eqnarray}
all the other commutators being equal to zero. According to equations (\ref{EL2}) and (\ref{MEL2})
we obtain
\begin{eqnarray}
B(x)&=& -\,i\int\mathrm d{\bf k}\;k_{\nu}\left[\,c_{\,{\bf k}\,S}\,\,v_{\,{\bf k}\,S}^{\,\nu}(x) -
c_{\,{\bf k}\,S}^{\,\dagger}\,v_{\,{\bf k}\,S}^{\,\nu\,\ast}(x)\,\right]_{k_{0}\,=\,\omega_{\,{\bf k}}}\\&=&m\int\mathrm d{\bf k}\;\left[\,b_{\,{\bf k}}\,u_{\,{\bf k}}(x) +
b_{\,{\bf k}}^{\,\dagger}\,u_{\,{\bf k}}^{\ast}(x)\,\right]_{k_{0}\,=\,\omega_{\,{\bf k}}}
\end{eqnarray}
in such a manner that the physical Hilbert space ${\mathfrak H}_{\,\rm phys}\,,$
with a positive semi-definite metric for all the MCS massive quantum states,
is selected out from the Fock space ${\mathfrak F}$ by means of the
customary subsidiary condition
\begin{eqnarray}
B^{\,(-)}(x)\,|\,\rm phys\,\rangle\;=\;0\qquad\quad
\forall\,|\,\rm phys\,\rangle\,\in\,{\mathfrak H}_{\,\rm phys}\subset{\mathfrak F}
\end{eqnarray}
On the other side,
the physical MCS massive quanta are created out of the Fock vacuum by the creation part of the quantum
physical massive MCS vector field $A^{\nu}_{\rm CS}(x)$,
with the standard non-vanishing commutation relations
\begin{displaymath}
 \left[\,c_{\,{\bf k},A}\,,\,c_{\,{\bf k}^{\prime},A^{\prime}}^{\,\dagger}\,\right]\;=\;\delta_{AA^{\prime}}\,
\delta({\bf k}-{\bf k}^{\prime})\qquad\qquad A,A^{\prime}=L,\pm
\end{displaymath}
all the other commutators being equal to zero.
Notice that the MCS massive 1-particle states
of definite spatial momentum $\bf k$ do exhibit three polarization states, i.e.,
one linear longitudinal polarization of real vector $\varepsilon^{\,\nu}_{\,L}(k)$
with dispersion relation $k^2=m^2$
and two chiral transverse states with complex vectors $\varepsilon^{\,\nu}_{\,\pm}(k_{\pm})$
and dispersion relations (\ref{MCSDR}),
the negative chirality states $\varepsilon^{\,\nu}_{\,-}(k_-)$ being well defined
only for $|\,\zeta_{\lambda}\cdot k_{-}^{\lambda}\,|<m/\zeta\,\Leftrightarrow\,k_{-}^{2}>0$.
\section{The Bogolyubov Transformations}

\bigskip
An equivalent rearrangement can be pursued in the right region R, where the
parity breaking medium is present. As a result we shall obtain the following normal modes expansion
of the Chern-Simons vector field for $ x\ge0\,: $ namely,
\begin{eqnarray}
A^{\nu}_{\rm CS}(\mathrm x)&=&
\int\mathrm d\hat{\rm k}\sum_{A=\pm,L}\,
\left[\,\mathrm c_{\,\hat{\rm k},\,A}\,v_{\,\hat{\rm k},\,A}^{\,\nu}(\mathrm x) +
\mathrm c_{\,\hat{\rm k},\,A}^{\,\dagger}\,v_{\,\hat{\rm k},\,A}^{\,\nu\,\ast}(\mathrm x)\,\right]
\end{eqnarray}
in which, in full analogy with the Proca-St\"uckelberg plane wave solution (\ref{PSprogressive}),
the incident or progressive plane wave solution for the
Chern-Simons classical field equations is provided by
\begin{eqnarray}
v_{\,\hat{\rm k}\,A}^{\,\nu}(\mathrm x) =\theta(x)\,
\left[\,(2\pi)^3\,2K_{A}\,\right]^{-\frac12}\,\varepsilon^{\,\nu}_{\,A}(\,\hat{\rm k})\
\exp\{\,i\,x K_{A}(\hat{\mathrm{k}}) - i\,\hat{\rm k}\cdot\hat{\rm x}\,\}\\
K_A=\left\lbrace
\begin{array}{cc}
\sqrt{\,\hat{\rm k}^2 - m^2\pm\sqrt{\zeta_{\,x}^{\,2}\,\hat{\rm k}^2}}
 &\quad {\rm for}\quad A=\pm\\
K=\sqrt{\,\hat{\rm k}^2 - m^2}&\quad {\rm for}\quad A=L
\end{array}\right.
\label{KKA}
\end{eqnarray}
while the canonical commutation relations still hold true, viz.,
\begin{eqnarray}
[\,{\mathrm c}_{\,{\hat{\rm k}},A}\,,\,{\mathrm c}_{\,\hat{\rm k}^{\,\prime}\!,\,B}\,]=
[\,{\mathrm c}^{\,\dagger}_{\,{\hat{\rm k}},A}\,,\,
{\mathrm c}^{\,\dagger}_{\,{\hat{\rm k}^{\,\prime}}\!,\,B}\,]=0
\qquad\quad
[\,{\mathrm c}_{\,{\hat{\rm k}},A}\,,\,{\mathrm c}^{\,\dagger}_{\,\hat{\rm k}^{\,\prime}\!,\,B}\,]=
\delta(\hat{\rm k}-\hat{\rm k}^{\,\prime})\,\delta_{\,AB}
\label{CCRsCS}
\end{eqnarray}
Coming back to the boundary conditions, it turns out that the matching condition for the vector fields
at the hyper-plane $ x=0 $ does take the form
\begin{eqnarray}
 &&\int\mathrm d\hat{\rm k}\left\lbrace \sum_{A}
\left[\,\mathrm c_{\,\hat{\rm k},A}\,v_{\,\hat{\rm k},A}^{\,\mu}(\hat{\rm x}) +
\mathrm c_{\,\hat{\rm k},A}^{\,\dagger}\,v_{\,\hat{\rm k},A}^{\,\mu\,\ast}(\hat{\rm x})\,\right]\right.\\&&\left.
- \sum_{r=1}^3\
\left[\,\mathrm a_{\,\hat{\rm k}\,,\,r}\,u^{\,\mu}_{\,\hat{\rm k},r}(\hat{\rm x})
+ \mathrm a^\dagger_{\,\hat{\rm k},\,r}\,
u^{\,\mu\,\ast}_{\,\hat{\rm k},r}(\hat{\rm x})\,\right]\right\rbrace =0\quad
\label{boundary3}
\end{eqnarray}
which can be implemented if we perform a Bogolyubov transformation
 \begin{equation}
 v_{\,\hat{\rm k},A}^{\,\nu}(\hat{\rm x})=\sum_{s=1}^3
 \,\left[\alpha_{s A}(\hat{\rm k})\,u^{\,\nu}_{\,\hat{\rm k},s}(\hat x) -
\beta_{s A}(\hat{\rm k})\,u^{\,\nu\,\ast}_{\,\hat{\rm k},s}(\hat{\rm x}) \right]
 \label{bg}
\end{equation}
where the complex numerical coefficients $\alpha_{s A}(\hat{\rm k})\,,\beta_{s A}(\hat{\rm k})$
will be determined below. On one hand we find
\begin{eqnarray}
\left\langle u^{\,\mu}_{\,\hat{\rm p},r}\,|\,v_{\,\hat{\rm k},A}^{\,\nu}\right\rangle
&\equiv& \int\mathrm d\hat{\rm x}\;\left.
u^{\,\mu\,\ast}_{\,\hat{\rm p},r}(x,\hat{\rm x})\,(-\,i)\!\buildrel \!\leftrightarrow\over {\partial_x}
v_{\,\hat{\rm k},A}^{\,\nu}(x,\hat{\rm x})\,\right\rfloor_{\,x\,=\,0}\nonumber\\
&=& \delta(\hat{\rm k}-\hat{\rm p})\,e^{\,\mu}_{\,r}(\hat{\rm k})\;
\frac{K_A+K}{2\surd{(K_{A}\,K)}}\;\varepsilon^{\,\nu}_{A}(\hat{\rm k})
\end{eqnarray}
and on the other hand we evidently obtain
\begin{eqnarray}
\left\langle u^{\,\mu}_{\,\hat{\rm p},r}\,|\,u_{\,\hat{\rm k},s}^{\,\lambda}\right\rangle
=\delta(\hat{\rm k} - \hat{\rm p})\,e^{\,\mu}_{\,r}(\hat{\rm k})\,e^{\,\lambda}_{\,s}(\hat{\rm k})
=\,-\,\left\langle u^{\,\mu\,\ast}_{\,\hat{\rm p},r}\,|\,u_{\,\hat{\rm k},s}^{\,\lambda\,\ast}\right\rangle;
\quad
\left\langle u^{\,\mu}_{\,\hat{\rm p},r}\,|\,u_{\,\hat{\rm k},s}^{\,\lambda\,\ast}\right\rangle=0
\end{eqnarray}
and thereby
\begin{eqnarray}
 &&\left\langle u^{\,\mu}_{\,\hat{\rm p},r}\,|\,v_{\,\hat{\rm k},A}^{\,\nu}\right\rangle
= \delta(\hat{\rm k}-\hat{\rm p})\,e^{\,\mu}_{\,r}(\hat{\rm k})\sum_{s=1}^3
\alpha_{s A}(\hat{\rm k})\,e^{\,\nu}_{\,s}(\hat{\rm k})
\\&&
\left\langle u^{\,\mu\,\ast}_{\,\hat{\rm p},r}\,|\,v_{\,\hat{\rm k},A}^{\,\nu}\right\rangle
= \delta(\hat{\rm k}-\hat{\rm p})\,e^{\,\mu}_{\,r}(\hat{\rm k})\sum_{s=1}^3
\beta_{s A}(\hat{\rm k})\,e^{\,\nu}_{\,s}(\hat{\rm k})
\end{eqnarray}
A comparison yields -- by omitting the argument $\hat{\rm k}  $
\begin{equation}
\frac{K_{A}+K}{2\surd{(K_{A}\,K)}}\;\varepsilon^{\,\nu}_{\,A}=
\sum_{s=1}^3 \alpha_{s A}\,e^{\,\nu}_{\,s}
\qquad\quad
\frac{K_{A}-K}{2\surd{(K_{A}\,K)}}\;\varepsilon^{\,\nu}_{\,A}=
\sum_{s=1}^3 \beta_{s A}\,e^{\,\nu}_{\,s}
\end{equation}
the solution of which is provided by
\begin{equation}
\alpha_{s A}=
{\textstyle\frac12}\,e_{\,s}\cdot\varepsilon_{A}\;
\frac{K_{A}+K}{\surd{(K_{A}\,K)}}
\qquad\quad
\beta_{s A}=
{\textstyle\frac12}\,e_{\,s}\cdot\varepsilon_{A}\;
\frac{K_{A}-K}{\surd{(K_{A}\,K)}}
\label{alphabeta}
\end{equation}
as it can be readily checked by direct inspection, where the wave vectors $ K,K_A $
are positive functions of $ \hat{\mathrm k} $ as given by eq.s (\ref{KKA}).
Let's evaluate the quantity
\begin{displaymath}
\sum_{s=1}^3\left[\, \alpha_{s A}({\hat k})\,\alpha_{s B}^{\,\ast}({\hat k})
-\beta_{s A}({\hat k})\,\beta^{\,\ast}_{s B}({\hat k})\,\right]
\qquad\qquad {\rm for}\quad A,B=L,\pm
\end{displaymath}
First we find
\begin{eqnarray*}
&&{\textstyle\frac12}\,g_{\mu\nu}\,\varepsilon^{\,\mu}_{A}(\hat{\rm k})\,
{\textstyle\frac12}\,g_{\rho\sigma}\,\varepsilon^{\,\rho\,\ast}_{B}(\hat{\rm k})
\sum_{s=1}^3 e^{\,\nu}_{\,s}(\hat{\rm k})\,\,e^{\,\sigma}_{\,s}(\hat{\rm k})\\
&=&{\textstyle\frac14}\,
\varepsilon^{\,\mu}_{A}(\hat{\rm k})\,\varepsilon^{\,\rho\,\ast}_{B}(\hat{\rm k})
\left( -\,g_{\rho\mu} + \frac{k_{\rho}k_{\mu}}{m^{2}}\right)
= -\,{\textstyle\frac14}\,g_{\rho\mu}\,
\varepsilon^{\,\mu}_{A}(\hat{\rm k})\,\varepsilon^{\,\rho\,\ast}_{B}(\hat{\rm k})
= {\textstyle\frac14}\,\delta_{AB}
\end{eqnarray*}
where use has been made of the transverse-like condition, as well as
the orthogonality relations, together with
the fact that the covariant linear polarization vectors have been chosen to be real.
Hence we eventually obtain
\begin{eqnarray*}
 \sum_{s=1}^3 \alpha_{s A}(\hat{\rm k})\,\alpha_{s B}^{\,\ast}(\hat{\rm k})=
\delta_{AB}\;\frac{(K_{A}+K)^2}
{4K_{A}\,K}\\
\sum_{s=1}^3 \beta_{s A}(\hat{\rm k})\,\beta^{\,\ast}_{s B}(\hat{\rm k})=
\delta_{AB}\;\frac{(K_{A}-K)^{2}}
{4K_{A}\,K}
\end{eqnarray*}
Subtraction of the above expressions yields the customary relation
\begin{equation}
\sum_{s=1}^3\left[\, \alpha_{s A}(\hat{\rm k})\,\alpha_{s B}^{\,\ast}(\hat{\rm k})
-\beta_{s A}(\hat{\rm k})\,\beta^{\,\ast}_{s B}(\hat{\rm k})\,\right]=\delta_{AB}
\qquad\qquad [\,A,B=L,\pm\,]
\label{Bogol1}
\end{equation}
and by making quite analogous manipulations one can readily check that the further
Bogolyubov relations
\begin{equation}
\sum_{s=1}^3\left[\, \alpha_{s A}(\hat{\rm k})\,\beta_{s B}^{\,\ast}(\hat{\rm k})
-\beta_{s A}(\hat{\rm k})\,\alpha^{\,\ast}_{s B}(\hat{\rm k})\,\right]=0
\qquad\qquad [\,A,B=L,\pm\,]
\label{Bogol2}
\end{equation}
\begin{equation}
\sum_{s=1}^3\left[\, \alpha_{s A}(\hat{\rm k})\,\beta_{s B}(\hat{\rm k})
-\beta_{s B}(\hat{\rm k})\,\alpha_{s A}(\hat{\rm k})\,\right]=0
\qquad\qquad [\,A,B=L,\pm\,]
\label{Bogol3}
\end{equation}

\medskip
Turning back to the boundary condition (\ref{boundary3}) and taking the Bogolyubov transformation
(\ref{bg}) into account, we can write the operator equality
\begin{eqnarray}
\label{relation}
\mathrm a_{\,\hat{\rm k},\,r}=\sum_{A=\pm,L}\left[ \,\alpha_{\,r A}(\hat{\rm k})\,\mathrm c_{\,\hat{\rm k},A}
- \beta^{\,\ast}_{\,r A}(\hat{\rm k})\,\mathrm c_{\,\hat{\rm k},A}^{\,\dagger}\,\right]
\label{yubov0} \\
\mathrm c_{\,\hat{\rm k},A}=\sum_{r=1}^3\left[ \,\alpha^{\,\ast}_{A r}(\hat{\rm k})\,\mathrm a_{\,\hat{\rm k},\,r}
+ \beta^{\,\ast}_{A r}(\hat{\rm k})\,\mathrm a_{\,\hat{\rm k},\,r}^{\,\dagger}\,\right]
\label{yubov1}
\end{eqnarray}
From the canonical commutation relations (\ref{CCRsCS}) we obtain
\begin{eqnarray*}
 \left[\,\mathrm a_{\,\hat{\rm k},\,r}\,,\,\mathrm a^{\dagger}_{\,\hat{\rm p},\,s}\,\right]
&=& \delta(\hat{\rm k}-\hat{\rm p})\,\delta_{\,rs}\\
&=&\sum_{A,B=\pm,L}\left[\,\alpha_{\,r A}(\hat{\rm k})\,\mathrm c_{\,\hat{\rm k},A}
- \beta^{\,\ast}_{\,r A}(\hat{\rm k})\,\mathrm c_{\,\hat{\rm k},A}^{\,\dagger}\,,\,
\alpha^{\,\ast}_{\,s B}(\hat{\rm p})\,\mathrm c^{\,\dagger}_{\,\hat{\rm p},B}
- \beta_{\,s B}(\hat{\rm p})\,\mathrm c_{\,\hat{\rm p},B}\,\right]\\
&=&\sum_{A=\pm,L}\left[\,\alpha_{\,r A}(\hat{\rm k})\,\alpha^{\,\ast}_{\,s A}(\hat{\rm p})
- \beta_{\,r A}(\hat{\rm k})\,\beta^{\,\ast}_{\,s A}(\hat{\rm p})\,\right]
\delta(\hat{\rm k}-\hat{\rm p})
\end{eqnarray*}
and consequently
\begin{equation}
 \sum_{A=\pm,L}\left[\, \alpha_{\,r A}(\hat{\rm k})\,\alpha^{\,\ast}_{\,s A}(\hat{\rm k})
- \beta_{\,r A}(\hat{\rm k})\,\beta^{\,\ast}_{\,s A}(\hat{\rm k})\,\right]=\delta_{\,rs}
\label{yubov2}
\end{equation}
Finally, the null commutators
$[ \,\mathrm a_{\,\hat{\rm k},\,r}\,,\,\mathrm a_{\,\hat{\rm p},\,s}\,]=
[ \,\mathrm a^{\dagger}_{\,\hat{\rm k},\,r}\,,\,\mathrm a^{\dagger}_{\,\hat{\rm p},\,s}\,]=0$
lead to the further relations
\begin{equation}
 \sum_{A=\pm,L}\left[\,\alpha_{\,r A}(\hat{\rm k})\,\beta^{\,\ast}_{\,s A}(\hat{\rm k})
- \beta_{\,r A}(\hat{\rm k})\,\alpha^{\,\ast}_{\,s A}(\hat{\rm k})\,\right]=0
\label{yubov3}
\end{equation}
\begin{equation}
 \sum_{A=\pm,L}\left[\,\alpha_{\,r A}(\hat{\rm k})\,\beta_{\,s A}(\hat{\rm k})
- \beta_{\,r A}(\hat{\rm k})\,\alpha_{\,s A}(\hat{\rm k})\,\right]=0
\label{yubov4}
\end{equation}
There are two different Fock vacuum states: namely,
\begin{displaymath}
\mathrm a_{\,\hat{\rm k},\,r}\vert0\rangle=0 \qquad\quad
\mathrm c_{\,\hat{\rm k},A}\mid\Omega\,\rangle=0
\end{displaymath}
whence from (\ref{yubov1})
\[\mathrm c_{\,\hat{\rm k},A}\mid0\,\rangle=
\sum_{r=1}^3\beta^{\,\ast}_{A r}(\hat{\rm k})\,\mathrm a_{\,\hat{\rm k},\,r}^{\,\dagger}\mid0\,\rangle\]
and consequently
\begin{eqnarray*}
 \langle\,0\mid \mathrm c_{\,\hat{\rm p},B}^{\,\dagger}\,\mathrm c_{\,\hat{\rm k},A}\mid0\,\rangle&=&
\sum_{r,s=1}^3\langle\,0\mid \mathrm a_{\,\hat{\rm p},s}\,\mathrm a_{\,\hat{\rm k},\,r}^{\,\dagger}\mid0\,\rangle\,
\beta_{B s}(\hat{\rm p})\,\beta^{\,\ast}_{A r}(\hat{\rm k})\\
&=&\delta(\hat{\rm k}-\hat{\rm p})\sum_{r=1}^3\beta_{B r}(\hat{\rm k})\,\beta^{\,\ast}_{A r}(\hat{\rm k})
=\delta(\hat{\rm k}-\hat{\rm p})\,\delta_{AB}\;\frac{(K_{A}-K)^{2}}
{4K_{A}\,K}
\end{eqnarray*}
In turn we evidently obtain
\begin{equation}
\left[ \, \mathrm a_{\,\hat{\rm k},\,r} +
\sum_{A=L,\pm}\beta^{\,\ast}_{\,r A}(\hat{\rm k})\,
\mathrm c_{\hat{\rm k},A}^{\,\dagger}\,\right] \mid\Omega\,\rangle=0
\end{equation}
that yields
\begin{eqnarray*}
 \langle\,\Omega\mid \mathrm a_{\,\hat{\rm p},s}^{\,\dagger}\,
 \mathrm a_{\,\hat{\rm k},r}\mid\Omega\,\rangle&=&
\sum_{A,B=L,\pm}\langle\,\Omega\mid \mathrm c_{\,\hat{\rm p},A}\,
\mathrm c_{\,\hat{\rm k},\,B}^{\,\dagger}\mid\Omega\,\rangle\,
\beta_{B s}(\hat{\rm p})\,\beta^{\,\ast}_{A r}(\hat{\rm k})\\
&=&\delta(\hat{\rm k}-\hat{\rm p})\sum_{A=L,\pm}\beta_{A s}(\hat{\rm k})\,
\beta^{\,\ast}_{A r}(\hat{\rm k})
\end{eqnarray*}
Moreover we get
\begin{eqnarray*}
 \langle\,0\mid \mathrm a_{\,\hat{\rm p},s}\,\mathrm c_{\,\hat{\rm k},A}\mid0\,\rangle
=\langle\,0\mid \mathrm a_{\,\hat{\rm p},s}\,\sum_{r=1}^3
\left[ \,\alpha^{\,\ast}_{A r}(\hat{\rm k})\,\mathrm a_{\,\hat{\rm k},\,r}
+ \beta^{\,\ast}_{A r}(\hat{\rm k})\,\mathrm a_{\,\hat{\rm k},\,r}^{\,\dagger}\,\right]\mid0\,\rangle\\
=\sum_{r=1}^3\beta^{\,\ast}_{A r}(\hat{\rm k})\,\langle\,0\mid
\mathrm a_{\,\hat{\rm p},s}\,\mathrm a^{\,\dagger}_{\,\hat{\rm k},\,r}\mid0\,\rangle
=\delta(\hat{\rm k}-\hat{\rm p})\,\beta^{\,\ast}_{A s}(\hat{\rm k})\\
\langle\,0\mid \mathrm a_{\,\hat{\rm p},s}\,\mathrm c^{\,\dagger}_{\,\hat{\rm k},A}\mid0\,\rangle
=\langle\,0\mid \mathrm a_{\,\hat{\rm p},s}\,
\sum_{r=1}^3\left[ \,\alpha_{A r}(\hat{\rm k})\,\mathrm a^{\,\dagger}_{\,\hat{\rm k},\,r}
+ \beta_{A r}(\hat{\rm k})\,\mathrm a_{\,\hat{\rm k},\,r}\,\right]\mid0\,\rangle\\
=\sum_{r=1}^3\alpha_{A r}(\hat{\rm k})\,\langle\,0\mid \mathrm a_{\,\hat{\rm p},s}\,
\mathrm a^{\,\dagger}_{\,\hat{\rm k},\,r}\mid0\,\rangle
=\delta(\hat{\rm k}-\hat{\rm p})\,\alpha_{A s}(\hat{\rm k})
\end{eqnarray*}
The latter quantity $\alpha_{A s}(\hat{\rm k})$ can thereof be interpreted as the
relative probability amplitude that a
particle of mass $m$, frequency  $\omega\,,$ wave vector $(K_{A}>0,k_y,k_z)$ and chiral polarization vector $\varepsilon^{\,\mu}_{A}(\hat{\rm k})$
is transmitted from the left face to the right face through the hyper-plane $ x=0$ by an incident
Proca-St\"uckelberg particle with equal mass $m$ and frequency $\omega\,,$ wave vector $(K>0, k_y, k_z)$
but polarization vector $e^{\,\mu}_{s}(\hat{\rm k})$.
As an effect of this transmission, the first component of wave vector of a
birefringent massive particle changes from $K$ to $K_\pm$, while the longitudinal massive
quanta do not change their wave vectors.
\section{The Functional Squeeze Operator Algebra}
In this Section we aim to apply the general method developed in
\cite{Soldati2011} to the present context, in order to calculate the squeezed pairs production
in the presence of a parity breaking medium. To start with,
consider e.g. the pair annihilation and production squeeze operators
for the Chern-Simons vector quanta
arising in the presence of a parity breaking background medium:
namely,
\begin{eqnarray}
\Pi(z)\,\equiv\,\sum_{\jmath\,\in\mathfrak Q}
{\textstyle\frac12}\,z_{\,\jmath}\,{\rm c}_{\,\jmath}^{\,2}\,=\,
\sum_{\jmath\,\in\mathfrak Q} {\textstyle\frac12}\,z_{\,\jmath}\,\Pi_{\,\jmath}\\
\Pi^{\,\dagger}(\bar z)\,\equiv\,\sum_{\imath\,\in\mathfrak Q}
{\textstyle\frac12}\,\bar z_{\,\imath}\,
{\rm c}_{\,\imath}^{\,\dagger\,2}\,=\,
\sum_{\imath\,\in\mathfrak Q}
{\textstyle\frac12}\,\bar z_{\,\imath}\,\Pi^{\,\dagger}_{\,\imath}
\end{eqnarray}
where we use the short notation
\[
\sum_{\imath\,\in\mathfrak Q}=\int{\mathrm d\hat{\rm k}}\sum_{A=\,\pm,L}
\qquad\quad(\,\hat{\rm k}\in{\mathbb R}^3\,)
\]
whereas $z_{\,\imath}\equiv z_{A}(\,\hat{\rm k}\,)$
are complex valued {\sf dimensionless} functions such that
\[
{\mathcal V}\sum_{\imath\,\in\mathfrak Q}\ z_{\,\imath}\,\bar z_{\,\imath}
=\mathcal{V}\int{\mathrm d\hat{\rm k}}\sum_{A=\,\pm,L}\vert\,
z_{A}(\,\hat{\rm k}\,)\,\vert^{2}\,=\,\nu_o
\]
is a {\sf pure number} that will be named the {\sf characteristic number}
of the given squeezed pairs distribution function
$z_{A}(\,\hat{\rm k}\,)\,,$ while ${\mathcal V}$ is the volume
of a symmetric and cubic box in the \textit{1+2}-dimensional  Minkowski space
$ \mathbb{M}^{2}_{1}\,, $
since we set as it is customary
\[
\lim_{\,\hat{k}\rightarrow0}\int\mathrm d\hat{x}\;\exp\left\lbrace \pm\, i\hat{k}\cdot\hat{x}\right\rbrace
=(2\pi)^3\,\delta(\hat{k}=0)=\lim_{\mathcal{V}\rightarrow\infty}
\int_{\mathcal{V}}\mathrm d\hat{x}\;\Longleftrightarrow\;
\delta(0)\doteqdot\frac{\mathcal{V}}{(2\pi)^{3}}
\]
It is worthwhile to remark that the whole algebraic construction we are going to set up in
the presence Section does actually live on the boundary hyper-plane $ \zeta\cdot x=0\,, $
as it is apparent from the above introduced notations in the related Fourier space.

The creation and destruction operators
for Chern-Simons vector particles $(\,{\rm c}_{\,\imath}\,,\,{\rm c}_{\,\jmath}^{\,\dagger}\,)$ and
Proca-Stueckelberg vector particles $(\,{\rm a}_{\,\imath}\,,\,{\rm a}^{\,\dagger}_{\,\jmath}\,)$
do satisfy e.g. the canonical commutation relations (\ref{CCRsCS})
\begin{equation}
[\,{\rm c}_{\,\imath}\,,\,{\rm c}^{\,\dagger}_{\,\jmath}\,]\,=\,\delta_{\,\imath\jmath}\,
=\,[\,{\rm a}_{\,\imath}\,,\,{\rm a}^{\,\dagger}_{\,\jmath}\,]
\qquad\quad(\,\imath,\jmath\,\in\mathfrak Q\,)
\end{equation}
all the remaining commutators being equal to zero. The Fock vacuum states are defined
in accordance with the Bogolyubov transformations (\ref{yubov0}) and (\ref{yubov1}),
that means
\begin{eqnarray}
&&\mathrm a_{\,\hat{\mathrm{k}},\,r}\,|\,0\,\rangle = 0 \quad
\Longleftrightarrow\quad
\sum_{A=\pm,L}\alpha_{\,r A}(\hat{\rm k})\,\mathrm c_{\,\hat{\rm k},A}\,|\,0\,\rangle
= \sum_{A=\pm,L}\beta^{\,\ast}_{\,r A}(\hat{\rm k})\,|\,\hat{\rm k}\ A\,\rangle\\&&
(\,\forall\,r=1,2,3\,\vee\, \hat{\mathrm{k}}\in\mathbb{R}^{3}\,)\nonumber
\label{vacuum}
\end{eqnarray}
\begin{eqnarray*}
&&\mathrm c_{\,\hat{\rm k},\,A}\,|\,\Omega\,\rangle = 0 \quad
\Longleftrightarrow\quad
\sum_{r=1}^3\left[\,\alpha^{\,\ast}_{A\,r}(\hat{\rm k})\,\mathrm a_{\,\hat{\rm k},\,r}
+ \beta^{\,\ast}_{A\,r}(\hat{\rm k})\,\mathrm a_{\,\hat{\rm k},\,r}^{\,\dagger}\,\right]
\,|\,\Omega\,\rangle = 0\\&&
(\,\forall\,A=\pm,L\,\vee\, \hat{\mathrm{k}}\in\mathbb{R}^{3}\,)
\end{eqnarray*}
If we denote by $Q_a\ (\,a=1,2,\ldots,n\,)$
any of the $n$ conserved charges of the system, which are allowed by the
parity breaking background field configuration, i.e.
\[
Q_a=\sum_{\jmath\,\in\mathfrak Q} q_{a\jmath}\,{\rm c}^{\,\dagger}_{\,\jmath}\,{\rm c}_{\,\jmath}
\qquad\quad(\,a=1,2,\ldots,n\,)
\]
where e.g. $q_{a\jmath}\ (\,a=1,2,\ldots,n\,)$ are the charges of the
squeezed state with definite quantum numbers $ \jmath\in\mathfrak Q\,, $
then for any squeezed state $\Pi_{\,\imath}^{\,\dagger}\,|\,0\,\rangle$
of definite quantum numbers we evidently find
\begin{equation}
Q_a\,\Pi_{\,\imath}^{\,\dagger}\,|\,\Omega\,\rangle=
2q_{a\imath}\,\Pi_{\,\imath}^{\,\dagger}\,|\,\Omega\,\rangle
\qquad\quad(\;\forall\,{\,\imath}\in{\mathfrak Q}\,\vee\,a=1,2,\ldots,n\,)
\end{equation}
The creation and annihilation squeeze operators satisfy the commutation relations
\begin{eqnarray}
\left[\,\Pi(z)\,,\,\Pi^{\,\dagger}(\bar z)\,\right]
=\;\sum_{\imath\,\in\mathfrak Q} {\textstyle\frac12}\,z_{\,\imath}\,\bar z_{\,\imath}
\left\{\mathrm c_{\,\imath}\,,\,{\rm c}^{\,\dagger}_{\,\imath}\right\}\ \equiv\ 2{\rm N}(\bar zz)
\end{eqnarray}
in which
\begin{eqnarray}
{\rm N}(\bar zz)
\,=\, {\rm N}^{\;\!\dagger}(z\bar z)
={\textstyle\frac14}\,\nu_o
+\sum_{\imath\,\in\mathfrak Q}  {\textstyle\frac12}\,z_{\,\imath}\,\bar z_{\,\imath}\,
{\rm c}^{\,\dagger}_{\,\imath}\,{\rm c}_{\,\imath}
\end{eqnarray}
Finally for $\nu_{\,\imath}\in{\mathbb R}\ (\,\forall\,\imath\,\in\mathfrak Q\,)$ we get
\begin{eqnarray}
\left[\,{\rm N}(\nu)\,,\,\Pi(z)\,\right]&=& -\sum_{\imath\,\in\mathfrak Q} {\textstyle\frac12}\,\nu_{\,\imath}\,
z_{\,\imath}\,\Pi_{\,\imath}\
=\ -\,\Pi(z\nu)\\
\left[\,{\rm N}(\nu)\,,\,\Pi^{\,\dagger}(\bar z)\,\right]&=&
\sum_{\imath\,\in\mathfrak Q} {\textstyle\frac12}\,\nu_{\,\imath}\,\bar z_{\,\imath}\,
\Pi^{\,\dagger}_{\,\imath}
=\ \Pi^{\,\dagger}(\nu\bar z)
\end{eqnarray}
It follows therefrom that the above three operators do satisfy the well known commutation relations
\begin{eqnarray}
&&\left[\,{\rm N}(\nu)\,,\,\Pi^{\,\dagger}(\bar z)\,\right]\ =\ \Pi^{\,\dagger}(\nu\bar z)\no
&&\left[\,\Pi(z)\,,\,{\rm N}(\nu)\,\right]\ =\ \Pi(\nu z)\label{su(2)}\\
&&\left[\,\Pi^{\,\dagger}(\bar z)\,,\,\Pi(z)\,\right]\ =\ 2{\rm N}(\bar zz)\nn
\end{eqnarray}
in which
\begin{eqnarray}
\Pi(z)=J_-(z)=J_1(z)-i\,J_2(z)\\
\Pi^{\,\dagger}(\bar z)=J_+(\bar z)=J_1(\bar z)+i\,J_2(\bar z)\\
{\rm N}(\nu)=J_3(z\bar z)
\end{eqnarray}
where the threesome of  operators
\begin{eqnarray}
J_1(z,\bar z)\equiv\textstyle\frac12\Big(\Pi(z)+\Pi^{\,\dagger}(\bar z)\Big)\no
J_2(z,\bar z)\equiv\frac{1}{2i}\,\Big(\Pi^{\,\dagger}(\bar z)-\Pi(z)\Big)\no
J_3(z\bar z)={\rm N}(\nu)
\label{angularmomenta}
\end{eqnarray}
are a basis of Hermitian generators obeying the well known SU(2) Lie algebra
\begin{eqnarray}
\left[\,J_a\,,\,J_b\,\right]=i\,\varepsilon_{\,abc}\,J_c\qquad\quad
(\,a,b,c=1,2,3\,)
\end{eqnarray}
Now it turns out that the quantum state $|\,\bar z\,\rangle\,,$ which represents a generic
squeezed pair of Chern-Simons particles with a
momentum distribution function $\bar z_{\,\imath}\ (\,\imath\,\in\mathfrak Q\,)\,,$ does satisfy
\begin{eqnarray}
|\,\bar z\,\rangle = \Pi^{\,\dagger}(\bar z)\,|\,\Omega\,\rangle
\qquad\quad
\langle\,z\,| = \langle\,\Omega\,|\,\Pi(z)
\end{eqnarray}
and exhibits the infrared regularized normalization
\begin{eqnarray}
\langle\,z\,|\,\bar z\,\rangle=\langle\,\Omega\,|\,[\,\Pi(z)\,,\Pi^{\,\dagger}(\bar z)\,]\,|\,\Omega\,\rangle
= -\,2\langle\,\Omega\,|\,{\rm N}(\bar zz)\,|\,\Omega\,\rangle = \nu_o
\end{eqnarray}
\subsection{Transmission and Reflection from the Algebraic Approach}
The general feature that characterizes the squeezed pairs production and annihilation
processes of massive vector particles in the presence of a parity breaking medium is the existence of
a non-singular Bogolyubov similarity transformation $\mathcal S\,,$
the generator of which is acting on the Fock space according to
\begin{eqnarray}
{\rm a}_{\,\imath}={\mathcal S}^{\,-1}\,{\rm c}_{\,\imath}\,{\mathcal S}
\equiv\sum_{\jmath\in\mathfrak Q}\Big(
\alpha_{\,\imath\jmath}\,{\rm c}_{\,\jmath}-\beta_{\,\imath\jmath}^{\,\ast}\,{\rm c}^{\,\dagger}_{\,\jmath}\Big)
\label{bogolyubov_bis}\\ \no
{\rm c}^{\,\dagger}_{\,\imath}={\mathcal S}^{\,-1}\,{\rm a}_{\,\imath}^{\,\dagger}\,{\mathcal S}
\equiv\,\sum_{\jmath\in\mathfrak Q}\Big(
\alpha_{\,\imath\jmath}\,{\rm a}^{\,\dagger}_{\,\jmath} + \beta_{\,\imath\jmath}\,{\rm a}_{\,\jmath}\Big)
\end{eqnarray}
in agreement with the previous relationships (\ref{yubov0}) and (\ref{yubov1}),
where
\begin{equation}
\begin{array}{c}
\alpha_{\,\imath\jmath}\equiv \alpha_{\,r A}(\hat{\rm k},\hat{\rm p})=
\alpha_{\,r A}(\hat{\rm k})\;\delta(\hat{\rm k}-\hat{\rm p})\,(2\pi)^3
\\ \\
\beta_{\,\imath\jmath}\equiv \beta_{\,r A}(\hat{\rm k},\hat{\rm p})=
\beta_{\,r A}(\hat{\rm k})\;\delta(\hat{\rm k}-\hat{\rm p})\,(2\pi)^3
\end{array}
\end{equation}
\begin{eqnarray}
\alpha_{\,r A}(\,\hat{\rm k}\,)
&=&{\textstyle\frac12}\,e_{\,r}(\hat{\rm k})\cdot\varepsilon_{A}(\hat{\rm k})\;\frac{K_{A}+K}{\surd(K_{A}K)}
\\
\beta_{\,r A}(\,\hat{\rm k}\,)
&=&{\textstyle\frac12}\,e_{\,r}(\hat{\rm k})\cdot\varepsilon_{A}(\hat{\rm k})\;\frac{K_{A}-K}{\surd(K_{A}K)}
\end{eqnarray}
together with
\[
\sum_{\imath\in\mathfrak Q}\left\lbrace
\begin{array}{c}
\alpha^{\,\ast}_{\,\imath\jmath}\,\alpha_{\,\imath\varkappa} -
\beta^{\,\ast}_{\,\imath\jmath}\,\beta_{\,\imath\varkappa} = \delta_{\,\jmath\varkappa}\\
\alpha_{\,\imath\jmath}\,\beta_{\,\imath\varkappa}^{\,\ast} -
\alpha^{\,\ast}_{\,\imath\jmath}\,\beta_{\,\imath\varkappa}=0
\end{array}
\right. \qquad\quad
(\,\forall\,\imath\,,\,\jmath\,,\,\varkappa\,\in\mathfrak Q\,)
\]
in such a manner that the canonical commutation relations (\ref{CCRsCS})
$$
[\,{\rm c}_{\,\imath}\,,\,{\rm c}^{\,\dagger}_{\,\jmath}\,]=\delta_{\imath\jmath}=
[\,{\rm a}_{\,\imath}\,,\,{\rm a}^{\,\dagger}_{\,\jmath}\,]\qquad\qquad{\rm et\ cetera}
$$
keep unchanged thanks to the similarity nature of the non-singular transformation $\mathcal S\,.$
Notice that in the present framework the Bogolyubov coefficients
$ \alpha_{\,\imath\jmath} $ and $ \beta_{\,\imath\jmath} $ can always be chosen to be real
-- see eq. (\ref{alphabeta}) for linear polarization.

It follows that we come to the two Fock spaces ${\mathfrak F}_{\,\rm PS}$ and
${\mathfrak F}_{\,\rm CS}$ which are generated by the cyclic vacuum states normalized to one and
defined by
\begin{eqnarray}
{\rm a}_{\,\imath}\,|\,0\,\rangle=0=\langle\,0\,\vert\,{\rm a}_{\,\imath}^{\,\dagger}
\quad\qquad
(\,\forall\,\imath\,\in\mathfrak Q\,)
\label{vacuumPS}\\
{\rm c}_{\,\jmath}\,|\,\Omega\,\rangle=0=\langle\,\Omega\,\vert\,{\rm c}_{\,\jmath}^{\,\dagger}
\quad\qquad
(\,\forall\,\jmath\,\in\mathfrak Q\,)
\label{vacuumCS}
\end{eqnarray}
Now we have, for example,
\begin{eqnarray}
{\rm c}_{\,\imath}\,{\rm a}^{\,\dagger}_{\,\varkappa}\,|\,0\,\rangle &=&\sum_{\jmath\in\mathfrak Q}
\Big(\alpha^{\,\ast}_{\,\imath\jmath}\,{\rm a}_{\,\jmath}\,{\rm a}^{\,\dagger}_{\,\varkappa}\,|\,0\,\rangle
+\beta_{\,\imath\jmath}^{\,\ast}\,
{\rm a}^{\,\dagger}_{\,\jmath}\,{\rm a}^{\,\dagger}_{\,\varkappa}\,|\,0\,\rangle\Big)\no
&=& \alpha^{\,\ast}_{\,\imath\varkappa}\,\,|\,0\,\rangle
+\sum_{\jmath\in\mathfrak Q}
\beta_{\,\imath\jmath}^{\,\ast}\,{\rm a}^{\,\dagger}_{\,\jmath}\,{\rm a}^{\,\dagger}_{\,\varkappa}\,|\,0\,\rangle
\end{eqnarray}
so that
\begin{eqnarray}
\langle\,0\,|\,{\rm c}_{\,\imath}\,{\rm a}^{\,\dagger}_{\,\jmath}\,|\,0\,\rangle
=\alpha^{\,\ast}_{\,\imath\jmath}\quad\qquad
(\,\forall\,\imath\,,\,\jmath\,\in\mathfrak Q\,)
\label{PSvacuum_persistence}
\end{eqnarray}
In a quite analogous way we get
\begin{eqnarray}
{\rm a}_{\,\imath}\,{\rm c}^{\,\dagger}_{\,\varkappa}\,|\,\Omega\,\rangle &=&\sum_{\jmath\in\mathfrak Q}
\Big(\alpha_{\,\imath\jmath}\,{\rm c}_{\,\jmath}\,{\rm c}^{\,\dagger}_{\,\varkappa}\,|\,\Omega\,\rangle
- \beta^{\,\ast}_{\,\imath\jmath}\,
{\rm c}^{\,\dagger}_{\,\jmath}\,{\rm c}^{\,\dagger}_{\,\varkappa}\,|\,\Omega\,\rangle\Big)\no
&=& \alpha_{\,\imath\varkappa}\,\,|\,\Omega\,\rangle
- \sum_{\jmath\in\mathfrak Q}
\beta^{\,\ast}_{\,\imath\jmath}\,{\rm c}^{\,\dagger}_{\,\jmath}\,{\rm c}^{\,\dagger}_{\,\varkappa}\,|\,\Omega\,\rangle
\end{eqnarray}
and thereby
\begin{eqnarray}
\langle\,\Omega\,|\,{\rm a}_{\,\imath}\,{\rm c}^{\,\dagger}_{\,\jmath}\,|\,\Omega\,\rangle
=\alpha_{\,\imath\jmath}\quad\qquad
(\,\forall\,\imath\,,\,\jmath\,\in\mathfrak Q\,)
\label{CSvacuum_persistence}
\end{eqnarray}
whence it follows that, as expected, the real and dimensionless  Bogolyubov coefficients
\[
\alpha_{\,r\,A}(\,\hat{\rm k}\,)=
{\textstyle\frac12}\,e_{\,r}(\hat{\rm k})\cdot\varepsilon_{A}(\hat{\rm k})\;\frac{K_{A}+K}{\surd(K_{A}K)}
\]
may be understood as the probability amplitudes  that a Proca-St\"{u}ckelberg pair of wave vector $ \hat{\rm k} $
is {\sc not created} out of the Fock vacuum $ |\,0\,\rangle $ or, in other words,
the \textsc{relative persistence probability amplitude} for the Proca-St\"{u}ckelberg vacuum.
In the very same manner the Bogolyubov coefficient
$\alpha_{\,r,A}(\hat{\rm k)}$ may be seen in turn as
the relative  persistence probability amplitude for the Chern-Simons vacuum $|\,\Omega\,\rangle$.
In a similar way, by taking the in vacuum expectation value
\begin{eqnarray}
\langle\,0\,|\,\Pi_{\,\imath}\,{\rm a}^{\,\dagger}_{\,\varkappa}\,{\rm c}_{\,\jmath}\,
|\,0\,\rangle
=\sum_{\ell\in\mathfrak Q}\beta_{\,\jmath\ell}^{\,\ast}\,
\langle\,0\,|\,{\textstyle\frac12}\,\mathrm a_{\,\imath}^{\,2}\,
{\rm a}^{\,\dagger}_{\,\ell}\,{\rm a}^{\,\dagger}_{\,\varkappa}\,|\,0\,\rangle
=\beta_{\,\jmath\imath}^{\,\ast}\;\delta_{\,\varkappa\imath}\;\,
(\,\forall\,\imath\,,\,\jmath\,,\,\varkappa\,\in\mathfrak Q\,)
\end{eqnarray}
it is also clear that we can understand the real Bogolyubov coefficient
\[
\beta_{\,r\,A}(\,\hat{\rm k}\,)=
{\textstyle\frac12}\,e_{\,r}(\hat{\rm k})\cdot\varepsilon_{A}(\hat{\rm k})\;\frac{K_{A}-K}{\surd(K_{A}K)}
\]
as the relative probability amplitude  that a  squeezed pair of Proca-St\"{u}ckelberg quanta
is created out or absorbed into the Proca-St\"{u}ckelberg vacuum.

\bigskip
To proceed further on, consider e.g. the Chern-Simons quantum vector field and let us define
\begin{eqnarray}
{\mathcal S}(\,\theta,\widehat{\bf n}\,)\;\equiv\;\exp\{-\,i\,\theta\cdot{\rm T}(z,\bar z,\nu)\}\\
\theta\cdot{\rm T}(z,\bar z,\nu)\equiv\,\Pi^{\,\dagger}(\,\theta\,\bar z)+\Pi(z\theta)+2{\rm N}(\theta\nu)
\end{eqnarray}
where $\theta_{\ell}\ (\,\forall\,\ell\,\in\mathfrak Q\,)$ is a real functional parameter,
while the functional unit vector
$\widehat{\bf n}_{\,\imath}$ is related to the functional parameters
$z_{\,\imath},\bar z_{\,\imath},\nu_{\,\imath}$ through the relationship
\[
\widehat{\bf n}_{\,\imath}^{\;\!2}=\nu_{\,\imath}^{\;\!2}-z_{\,\imath}\,\bar z_{\,\imath}=1\qquad\quad
(\,\forall\,\imath\,\in\mathfrak Q\,)
\]
For example, a suitable functional parametric form is provided by a set of
hyperbolic and trigonometric variables $(\,\upsilon_{\,\imath},\phi_{\,\imath}\,)\,,$
in such a manner that
\[
\nu_{\,\imath}=\cosh\upsilon_{\,\imath}\qquad\quad
z_{\,\imath}=\sinh\upsilon_{\,\imath}\,\exp\{-\,i\phi_{\,\imath}\}\qquad\quad
(\,\upsilon_{\,\imath}\in\mathbb R\,,\,0\le\phi_{\,\imath}<2\pi\,,\,\forall\,\imath\in\mathfrak Q\,)
\]
From the basic commutation relation
\begin{eqnarray}
[\,{\rm T}(z,\bar z,\nu)\,,\,{\rm c}_{\,\imath}\,]=
-\,\bar z_{\,\imath}\,{\rm c}_{\,\imath}^{\,\dagger}-\nu_{\,\imath}\,{\rm c}_{\,\imath}\qquad
[\,{\rm T}(z,\bar z,\nu)\,,\,{\rm c}^{\,\dagger}_{\,\imath}\,]= z_{\,\imath}\,{\rm c}_{\,\imath}
+ \nu_{\,\imath}\,{\rm c}^{\,\dagger}_{\,\imath}
\end{eqnarray}
we readily find
\begin{equation}
[\,{\rm T}\,,\,[\,{\rm T}\,,\,{\rm c}_{\,\imath}\,]\,]={\rm c}_{\,\imath}\qquad
[\,{\rm T}\,,\,[\,{\rm T}\,,\,{\rm c}^{\,\dagger}_{\,\imath}\,]\,]={\rm c}^{\,\dagger}_{\,\imath}
\end{equation}
As a consequence, we actually obtain the general Bogolyubov transformations in the form
\begin{eqnarray}
{\rm a}_{\,\imath}&=&(\cos\theta_{\,\imath} -i\,\nu_{\,\imath}\,\sin\theta_{\,\imath})\,{\rm c}_{\,\imath}\,-\,
i\,{\rm c}_{\,\imath}^{\,\dagger}\,\bar z_{\,\imath}\sin\theta_{\,\imath}
=\;\alpha_{\,\imath}\,{\rm c}_{\,\imath}\;-\;\beta_{\,\imath}^\ast\,{\rm c}^{\,\dagger}_{\,\imath}\\
{\rm a}^{\,\dagger}_{\,\jmath}&=&\left(\cos\theta_{\jmath}+i\,\nu_{\,\jmath}\,
\sin\theta_{\jmath}\right)\,{\rm c}^{\,\dagger}_{\,\jmath}\,+\,i\,{\rm c}_{\,\jmath}\,
z_{\,\jmath}\sin\theta_{\jmath}
=\;\alpha_{\,\jmath}^\ast\,{\rm c}^{\,\dagger}_{\,\jmath}\;-\;\beta_{\,\jmath}\,{\rm c}_{\,\jmath}
\end{eqnarray}
with
\begin{equation}
\alpha_{\,\jmath}\;\equiv\;\cos\theta_{\jmath} -i\,\nu_{\,\jmath}\,\sin\theta_{\jmath}\qquad\quad
\beta_{\,\jmath}^{\,\ast}\;\equiv\;i\,\bar z_{\,\jmath}\sin\theta_{\jmath}
\end{equation}
\[
\widehat{\bf n}_\jmath^{\;\!2}\,=\,
\nu_\jmath^{\;\!2} - \bar z_\jmath\,z_\jmath\;=\;1\qquad\quad
|\,\alpha_{\,\jmath}\,|^{\;\!2}\;+\;|\,\beta_{\,\jmath}\,|^{\;\!2}\ =\ 1
\]
It follows thereby that the functional unitary operator
\begin{equation}
{\mathcal S}(\theta,z,\nu)={\mathcal S}(\,\theta,\widehat{\bf n}\,)=
{\mathcal S}(\,\alpha,\beta\,)\qquad\quad{\mathcal S}^{\,-1}={\mathcal S}^{\,\dagger}
\end{equation}
does generate the Bogolyubov similarity transformations which connect the Proca-St\"uckelberg and
Chern-Simons vector fields on the boundary hyper-plane $ \zeta\cdot x=0 $
according to the suitable definitions
\begin{eqnarray}
A^{\nu}_{\,\rm PS}(\hat{\mathrm{x}})\;=\;
{\mathcal S}^{\,-1}\,A^{\nu}_{\,\rm CS}(\hat{\mathrm{x}})\,{\mathcal S}
\qquad\quad\qquad
|\,{\rm CS}\,\rangle\;=\;{\mathcal S}\,|\,{\rm PS}\,\rangle
\end{eqnarray}
Moreover we obtain
\begin{eqnarray}
A^{\nu}_{\,\rm PS}(\hat{\mathrm{x}})&=&
\sum_{\imath\,\in\mathfrak Q}\;{\mathcal S}^{\,-1}
\left[\,{\rm c}_{\,\imath}\;v_{\,\imath}(\hat{\mathrm{x}}) +\
{\rm c}^{\,\dagger}_{\,\imath}\;v^{\,\ast}_{\,\imath}(\hat{\mathrm{x}})\,\right]\,
{\mathcal S}\nonumber\\
&=&\sum_{\jmath\,\in\mathfrak Q}\;
\left[\,{\rm a}_{\,\jmath}\;u_{\,\jmath}(\hat{\mathrm{x}}) +\
{\rm a}^{\,\dagger}_{\,\jmath}\;u^{\,\ast}_{\,\jmath}(\hat{\mathrm{x}})\,\right]\\
A^{\nu}_{\,\rm CS}(\hat{\mathrm{x}})&=&
\sum_{\imath\,\in\mathfrak Q}\;{\mathcal S}
\left[\,{\rm a}_{\,\imath}\;u_{\,\imath}(\hat{\mathrm{x}}) +\
{\rm a}^{\,\dagger}_{\,\imath}\;u^{\,\ast}_{\,\imath}(\hat{\mathrm{x}})\,\right]\,
{\mathcal S}^{\,-1}\nonumber\\
&=&\sum_{\jmath\,\in\mathfrak Q}\;
\left[\,{\rm c}_{\,\jmath}\;v_{\,\jmath}(\hat{\mathrm{x}}) +\
{\rm c}^{\,\dagger}_{\,\jmath}\;v^{\,\ast}_{\,\jmath}(\hat{\mathrm{x}})\,\right]
\end{eqnarray}
where, in the case of the parity breaking medium and Minkowski space in four
dimensions, we have e.g.
\[
u_{\,\imath}(\hat{\mathrm{x}})\equiv u_{\,\hat{\bf k},r}^{\,\nu}\;\!(x=0,\hat{\rm x})
\qquad\quad
v_{\,\jmath}(\hat{\mathrm{x}})\equiv v_{\,\hat{\bf p},A}^{\,\nu}\;\!(x=0,\hat{\rm x})
\]
Hence, the general Bogolyubov transformation is nothing but
a functional rotation in the Fock space with parameter functions
$(\theta,z,\nu)=(\theta,\widehat{\bf n})=
(\,\alpha,\beta\,)\,,$ the generators of which are the squeezed pairs emission $\Pi^{\,\dagger}(\theta\bar z)\,,$
the squeezed pairs absorption $\Pi(\theta z)$ and the squeezed pairs number ${\rm N}(\theta\nu)$ operators,
which actually fulfill the functional commutation relations arising from the SU(2) Lie algebra.
\section{The extended Sauter-Schwinger-Nikishov Formula}
Suppose that at some initial time ${\rm t}_{\imath}$ our system is in a definite state,
e.g. the vacuum $|\,0\;{\rm in}\,\rangle$ as defined in equation (\ref{vacuum}),
so that it does not contain
any Proca-St\"{u}ckelberg
vector particles, and that particle detectors are homogeneously distributed and placed within
a small but finite spatial volume $ \Delta\upsilon $.
Then the final state at some later although close time ${\rm t}_{f}$, with $ t_{f}-t_{\imath}=\Delta t $,
has some calculable probability to contain zero, one, two, etc. emitted squeezed pairs of vector particles
of the Proca-St\"{u}ckelberg or Chern-Simons particles.
For example the LIV vacuum persistence probability amplitude,
i.e., the probability amplitude of no emission of squeezed Chern-Simons pairs at the time ${\rm t}_{f}$,
will be given by
\begin{equation}
\langle\,\Omega\;\mathrm{out}\,|\,0\;\mathrm{in}\,\rangle=
\langle\,0\;{\rm out}\,|\,{\mathcal S}^{\,\dagger}\,|\,0\;{\rm in}\,\rangle
\end{equation}
For this interpretation to make sense, one has to actually verify that the vacuum to vacuum
transition probability
$$
W_{0\,,\,f\,\leftarrow\,\imath}\;\equiv\;
|\,\langle\,\Omega\;\mathrm{out}\,|\,0\;\mathrm{in}\,\rangle\,|^{\;\!2}
$$
is not greater than one. To this concern consider the Hermitean operators which correspond to the
number of CS and PS quanta with quantum numbers $ \imath\in\mathfrak Q $: namely,
\begin{eqnarray*}
\mathrm N^{\,\rm CS}_{\,\imath}\;&\equiv&\;{\textstyle\frac12}
\left(\,{\rm c}^{\,\dagger}_{\,\imath}\,{\rm c}_{\,\imath} +
{\rm c}_{\,\imath}\,{\rm c}^{\,\dagger}_{\,\imath}\,\right)\,=\,
{\mathcal S}\,\mathrm N^{\,\rm PS}_{\,\imath}\,{\mathcal S}^{\,-1} =
 \mathrm N^{\,\rm PS}_{\,\imath}\;+\;\alpha_{\,\imath}\,\beta_{\,\imath}^{\,\ast}\,\Pi_{\,\imath}^{\,\dagger}
\;+\;\alpha^{\,\ast}_{\,\imath}\,\beta_{\,\imath}\,\Pi_{\,\imath}
\end{eqnarray*}
where
\[
\Pi_{\,\imath}\equiv {\rm a}_{\,\imath}^{\,2}\qquad\quad
4\,|\,\alpha_{\,\jmath}\,\beta^{\,\ast}_{\,\jmath}\,|^{\;\!2} +
\left(\,|\,\alpha_{\,\jmath}\,|^{\,2}-|\,\beta_{\,\jmath}\,|^{\,2}\,\right)^2=1
\]
It follows that if we set
\begin{equation}
\xi_{\,\jmath}\;\equiv\;2\,\alpha_{\,\jmath}\,\beta_{\,\jmath}^{\,\ast}\;,\qquad\quad
w_\jmath\;\equiv\;|\,\alpha_{\,\jmath}\,|^{\,2} - |\,\beta_{\,\jmath}\,|^{\,2}
\end{equation}
we can write the following {\sf squeeze pair operator Golden Rule} \cite{Soldati2011} that actually occurs
$\forall\,\imath,\jmath,\ell,\ldots\,\in\mathfrak Q\,:$
\begin{equation}
2\,\mathrm N^{\,\rm CS}_{\imath}\;=\;\xi_{\,\imath}\,\Pi_{\,\imath} +
\bar\xi_{\,\imath}\,\Pi^{\,\dagger}_{\,\imath} + 2w_{\,\imath}\,\mathrm N^{\,\rm PS}_{\imath}
\qquad\quad
\bar\xi_{\,\imath}\,\xi_{\,\imath} + w_{\,\imath}^{\;\!2}=1
\label{golden}
\end{equation}
It is important to remark that the above equality (\ref{golden})
holds true thanks to the similarity property satisfied by the
Bogolyubov coefficients $\alpha_{\,\imath}\,,\,\beta_{\,\jmath}\,.$
Then, {\bf for any complex distribution function} $\varphi\;\!(\hat{\mathrm{k}};A,r)
=\varphi_{\,\imath}\,,$ we can write
\begin{eqnarray}
2\,{\mathrm N}^{\,\rm CS}(\varphi) &\equiv&
\sum_{\imath\,\in\mathfrak Q}\;2\mathrm N^{\,\rm CS}_{\,\imath}\,\varphi_{\,\imath}\,=\,
\sum_{\imath\,\in\mathfrak Q}\;
\{{\rm c}^{\,\dagger}_{\,\imath}\,,\,{\rm c}_{\,\imath}\}\;\varphi_{\,\imath}
\nonumber\\
&=&\Pi^{\,\dagger}(\bar{\xi}\varphi)+\Pi(\xi\varphi)+2\mathrm N^{\,\rm PS}(w\varphi)\,=\,
\varphi\,\cdot\,{\rm T}(\xi,\bar{\xi},w)
\end{eqnarray}
In the application of the above general setting to the present circumstance, it should be kept in mind
that, owing to the presence of the boundary $ \zeta\cdot\mathrm{x}=0 $, then translation invariance
holds true only in the time evolution and in the transverse foliation $ \zeta\cdot\rm x= constant\,.$
It follows thereby that we can write
\begin{eqnarray}
2\,\mathrm N^{\,\rm CS}_{\imath}\,|\,0\;{\rm in}\,\rangle
&=&[\,{\rm a}_{\,\imath}\,,\,{\rm a}^{\,\dagger}_{\,\imath}\,]\,|\,0\;{\rm in}\,\rangle
=\delta_{\imath\imath}\,|\,0\;{\rm in}\,\rangle\nonumber\\
&=&\delta^{\,(3)}({0})\,|\,0\;{\rm in}\,\rangle
\equiv\,\Delta\sigma\Delta t\,(2\pi)^{-\,3}\,|\,0\;{\rm in}\,\rangle
\end{eqnarray}
where $\Delta\sigma$ denotes the small unit area in the $Oyz-$plane - e.g. a section of a region
in the $3-$dimensional space where the particle detectors are placed. Thus we eventually obtain
\begin{eqnarray}
\langle\,\Omega\;\mathrm{out}\,|\,0\;\mathrm{in}\,\rangle
&=&\langle\,0\;{\rm out}\,|\,{\mathcal S}_{\,\varphi}^{\,-1}\,|\,0\;{\rm in}\,\rangle\nonumber\\
&=&\langle\,0\;{\rm out}\,|
\exp\left\{\,2i\,{\mathrm N}^{\,\rm CS}(\varphi)\right\}|\,0\;{\rm in}\,\rangle
\nonumber\\
&=&\prod_{\,\imath\,\in\mathfrak Q}\;
\exp\left\lbrace\,i\,\varphi_{\,\imath}\,\Delta\sigma\Delta t\,(2\pi)^{-\,3}\right\rbrace
\nonumber\\
&=&\exp\left\lbrace\,i\,\Delta\sigma\Delta t\,(2\pi)^{-\,3}
\textstyle\sum_{\,\imath\,\in\mathfrak Q}\;\varphi_{\,\imath}\,\right\rbrace
\end{eqnarray}
However, according to the natural interpretation
which arises from eq. (\ref{PSvacuum_persistence}),
it is mandatory for consistency to identify
\begin{equation}
\varphi_{\,\jmath}\,\equiv\,
i\ln\,\alpha_{\,\jmath}^{\,\ast}
={\rm Arg}\,\alpha_{\,\jmath}+{\textstyle\frac12}\,i\ln\,|\,\alpha_{\,\jmath}\,|^{\,2}
\qquad\quad
(\,\forall\,\jmath\,\in\,{\mathfrak Q}\,)
\end{equation}
As a matter of fact, according to Nikishov \cite{nikishov}, the logarithm of
vacuum to vacuum transition amplitude is provided by
\begin{equation}
\ln\,\langle\,\Omega\;\mathrm{out}\,|\,0\;\mathrm{in}\,\rangle =
-\,\Delta\sigma\Delta t\,(2\pi)^{-\,3}
\sum_{\,\imath\,\in\mathfrak Q}\;\ln \alpha_{\,\imath}^{\,\ast}
\end{equation}
As a consequence it is possible to express the out vacuum in terms of the in operators
in the explicit form
\begin{eqnarray}
\langle\,\Omega\;\mathrm{out}\,| &=& \langle\,0\;\mathrm{out}\,|\,{\mathcal S}_{\,\varphi}^{\,-1}
\nonumber\\
&=&\langle\,0\;\mathrm{out}\,|\,\exp\left\{2i\,{\mathrm N}^{\,\rm CS}(\varphi)\right\}
\nonumber\\
&=& \langle\,0\;\mathrm{out}\,|\,\exp\left\{-\,2\,{\mathrm N}^{\,\rm CS}(\,\ln\,\alpha^{\,\ast}\,)\right\}
\nonumber\\
&=&\langle\,0\;\mathrm{out}\,|\,\exp\left\{-\textstyle\sum_{\imath\,\in\mathfrak Q}\;
{\rm c}^{\,\dagger}_{\,\imath}\,{\rm c}_{\,\imath}\,\ln\,\alpha_{\imath}^{\,\ast}\right\}
\end{eqnarray}
whence it immediately follows that, by the very construction,
\begin{equation}
{\rm c}_{\,\varkappa}\,|\,0\;{\rm out}\,\rangle\,=\,0\qquad\quad
(\,\forall\,\varkappa\,\in\,{\mathfrak Q}\,)
\end{equation}
Notice that the non-singular operator ${\mathcal S}_{\,\varphi}$
is not unitary, owing to the presence of an imaginary part in
the distribution function $\varphi=i\ln\,\alpha^{\,\ast}\,.$
Furthermore, from the golden operator identity (\ref{golden})
it follows that we can write
\begin{equation}
2\,{\mathcal N}(\varphi)=
\varphi\cdot{\rm T}(\xi,\bar{\xi},w)\equiv\,
\Pi^{\,\dagger}(\,\varphi\,\bar{\xi})+\Pi(\xi\varphi)+2{\rm N}(\varphi w)
\end{equation}
in such a manner that the out vacuum state can be expressed {\em \`a la} Dirac
as an infinite sea of squeezed pairs, i.e., a coherent like state involving any number
of squeezed pairs of any quantum numbers: namely,
\begin{eqnarray}
&&\langle\,\Omega\;\mathrm{out}\,|\,=
\langle\,0\;\mathrm{out}\,|\,{\mathcal S}_{\,\varphi}^{\,-1}\,=
\langle\,0\;\mathrm{out}\,|\,\exp\left\{-\,2\;{\mathcal N}(\,\ln\,\alpha^{\,\ast}\,)\right\}
\nonumber\\
&& = \langle\,0\;\mathrm{out}\,|\,\exp\left\{-\,\Pi^{\,\dagger}(\,\bar{\xi}\ln\,\alpha^{\,\ast}\,)-\,
\Pi(\,\xi\,\ln\,\alpha^{\,\ast}\,)-\,2{\rm N}(\,w\,\ln\,\alpha^{\,\ast}\,)\right\}
\end{eqnarray}
in which
\[
\xi_{\,\jmath}\;\equiv\;2\,\alpha_{\,\jmath}\,\beta^{\,\ast}_{\,\jmath}\qquad\quad
w_\jmath\;\equiv\;|\,\beta_{\,\jmath}\,|^{\,2} - |\,\alpha_{\,\jmath}\,|^{\,2}
\qquad\quad(\,\forall\,\jmath\in{\mathfrak Q}\,)
\]
Finally, one can readily generalize to the present circumstance the case of spinor QED
in the presence of a uniform electric field on the four dimensional Minkowski space: actually we have
\begin{equation}
\langle\,\Omega\;{\rm out}\,|\,0\;{\rm in}\,\rangle=
\exp\left\lbrace -\,i\,\Delta\sigma\Delta t\,(2\pi)^{-3}\sum_{A}
\int\mathrm{d}\hat{k}\;\varphi\;\!(\alpha)\right\rbrace
\end{equation}
with
\[
\hat{k}\,\equiv\,(\,\omega,k_y,k_z\,)
\qquad\quad
\alpha(\hat{k},A)\,\equiv\,{\textstyle\frac12}\cdot\frac{K_{A}+K}{\surd(KK_{A})}
\qquad\quad(\,A=\pm,L\,)
\]
where $ K $ and $ K_{A} $ are provided by equation (\ref{KKA}),
in such a manner that we can eventually  write
\begin{eqnarray}
|\,\langle\,\Omega\;{\rm in}\,|\,0\;{\rm out}\,\rangle\,|^{\;\!2}\;
&=& \exp\left\lbrace -\,\Delta\sigma\Delta t\, (2\pi)^{-3}\sum_{A}\int\mathrm{d}\hat{\mathrm{k}}\;
\ln\,|\,\alpha(\hat{\mathrm{k}},A)\,|^{\;\!2}
\right\rbrace\nonumber\\
&=&\exp\left\lbrace -\,2\Delta\sigma\Delta t\,(2\pi)^{-3}
\sum_{A=\,\pm}\int\mathrm{d}\hat{\mathrm{k}}\;\ln\alpha(\hat{\mathrm{k}},A)\right\rbrace
\end{eqnarray}
which is nothing but the suitable generalization of celebrated Sauter-Schwinger formula
\cite{SS} to the present context of a transition across a parity breaking medium from the Minkowski space and \emph{viceversa}. For example, to the leading order in the LIV relevant
scales  $ \zeta\ll m\ll M $, which properly characterize the present framework,
explicit evaluation shows  -- see Appendix  --
that the logarithm of the vacuum persistence probability trough the boundary per unit time and unit transverse area is actually approximately given  by
\begin{equation}
\frac{-\,1}{\Delta t\Delta\sigma}\,\ln |\,\langle\,\Omega\;{\rm in}\,|\,0\;{\rm out}\,\rangle\,|
\approx
\zeta\;\frac{mMc^{\,3}}{16\pi^{2}\hslash^{2}}\left\lbrace
1 + \frac{\hslash\zeta}{2mc}\left( 2\ln\frac{M}{m} -
\ln\frac{\hslash\zeta}{2mc}\right) + O\left(\frac{\hslash\zeta}{mc}\right)\right\rbrace
\end{equation}
Notice that, as expected, we recover the null result in the Lorentz invariant limit
$ \zeta\rightarrow0 $ when a unique cyclic vacuum state is left in the Fock space,
while the vacuum persistence probability becomes not completely negligible for\\
$ \zeta\Delta t\Delta\sigma\,mMc^{\,3}\simeq 8\pi^{2}\hslash^{2}\,.$
\section{Conclusion}
The problem of propagation of vector particles through a boundary between a parity breaking medium and an empty vacuum is quite interesting as it may arise in heavy ion and astro- physics. In this paper we restricted ourselves with the Chern-Simons dynamics generated by a CS vector orthogonal to the flat boundary which perfectly guaranties
the gauge invariance while provides the Lorentz symmetry violation in the Maxwell-Chern-Simons part coherent with
boundary implementation. On the other half-space we have a massive quantum vector field in vacuum. We investigated the different quantum field aspects of the model described before at quasi-classical level. In particular, the transmission/reflection coefficients for different cases were found in \cite{AKS11,Andrianov:2013bxa} and the possible influence on the cooling rate of neutron star was shown in \cite{aaek2016}.

However, the results of the present work allow us to describe the particle transition through the boundary between areas with different field equations at quantum level in a more general form. The fact that Bogolyubov transformations may be rewritten as a functional rotation in the Fock space is useful in problems not only connected with MCS electrodynamics in a finite volume, but in other investigations concerned with different Fock spaces for geometrically and physically different space-time regions and propagation between them.

We draw attention of a reader also to the probability of inducing finite volume bubbles of distinct matter/radiation background undertaken in the Section 5 and in the Appendix. It looks useful for description of phase transition to and from the parity breaking media in heavy ion collisions. However we stress that for infinite volume subspaces the total transition probability vanishes and the two media coexist. This is justified by calculation
of the probability to find any number of squeezed pairs of vector particles of the Proca-St¨uckelberg or Chern-Simons
kind in crossing the boundary between the empty space-time and the parity breaking medium.

\section*{Acknowledgements}
A.A. and S.K.
were supported by Grant RFBR projects 16-02-00348 and also got a
financial support of SPbSU.
A.A. was also supported through grants
FPA2013-46570, 2014-SGR-104 and Consolider CPAN. As well funding was
partially provided by the Spanish MINECO under project MDM-2014-0369
of ICCUB (Unidad de Excelencia `Maria de Maeztu').
This collaboration has been partially supported by grants from I.N.F.N. Sezione di Bologna
Inviti 2015 and RFO 2014  DIFA Universit\'a di Bologna.

%
\section*{Appendix A: Probability Flux}
In this Appendix we will calculate, to the leading approximation in the relevant physical
parameters of the present model, the probability flux from the Proca-St\"uckelberg vacuum state
$ |0\rangle $ to the Chern-Simons vacuum  state $ |\Omega\rangle $ and \emph{viceversa},
due to transmission and reflection of squeezed pairs quanta through the boundary that
separates the parity breaking medium from the Minkowski empty space.
To start with, we have to invert the Chern-Simons frequency formula for $ \hat{\mathrm{k}}^2>m^2 $
and $ \zeta^{\lambda}=(0,\zeta,0,0)\ (\,\zeta>0\,) $: we find
\begin{eqnarray*}
&&\omega_{\,{\bf k}\,,\,\pm}=
\sqrt{{\bf k}^2+m^{2}+\textstyle{\frac12}\zeta^{\,2}\pm\zeta\sqrt{k_x^2+m^2+\frac14\zeta^{\,2}}}
\\
&&\textstyle\hat{\mathrm{k}}^{2}-K^{2}-m^{2}-\frac12\zeta^{\,2}=
\zeta\,\sqrt{K^2+m^2+\frac14\zeta^{\,2}}
\ \Longleftrightarrow \
\{K^2-(\,\hat{\mathrm{k}}^2-m^2)\}^2-\hat{\mathrm{k}}^{2}\zeta^{\,2}=0
\\
&&K^2-(\,\hat{\mathrm{k}}^2-m^2)=\pm\sqrt{\hat{\mathrm{k}}^{2}\zeta^{\,2}}
\, \Longleftrightarrow \, K_{\pm}(\hat{\mathrm{k}})=
\sqrt{\hat{\mathrm{k}}^2-m^2\pm\zeta\sqrt{\hat{\mathrm{k}}^{2}}}
\end{eqnarray*}
As we have already emphasized the general requirements
$ k_{-}^2> 0\,\vee\,K_{-}(\hat{\mathrm{k}})> 0 $  entail the  conservative
 lower and upper approximate bounds for angular frequency
\[ m(1+\frac{\epsilon}{2}) \lesssim\omega\lesssim M\qquad\qquad
\epsilon=\frac{\hslash\zeta}{mc}
\]
where, as already mentioned, the UV cut-off $ M $ is the natural albeit model-dependent
axion-like decay constant size, typically ranging within $ 10^{9} \div10^{17}$ GeV,
according to the recent estimates - see e.g. the review paper
\cite{axionreview}.
If we suppose here that e.g. $ \zeta\ll m $, then  we have for $ \epsilon=(\zeta/m)\ll1 $
\begin{eqnarray*}
&&K(\,\hat{\mathrm{k}}\,)=\sqrt{\,\hat{\rm k}^2 - m^2}>0\qquad\quad
\mathrm{for}\quad \hat{\rm k}^2 > m^2\\
&&K_{\pm}(\,\hat{\mathrm{k}}\,)=K(\hat{\mathrm{k}})\,
\sqrt{\,1\pm\chi(\hat{\mathrm{k}})}>0\qquad\quad
\mathrm{for}\quad\hat{\rm k}^2 > m^2\left[\,
1+\epsilon^{2}\left(1\pm\sqrt{\frac{1}{\epsilon^{2}}+\frac14}\;\right)\right]
\end{eqnarray*}
\[
\chi(\,\hat{\mathrm{k}}\,)=
\frac{\zeta\sqrt{\hat{\mathrm{k}}^2}}{\hat{\rm k}^2 - m^2} \lesssim 1
\qquad\quad\mathrm{for}\quad\hat{\mathrm{k}}^2\gtrsim m^2(1+\epsilon)
\]
Notice that this particular function $ \chi(\hat{\mathrm{k}}) $ is a monotonically
decreasing function for $ \hat{\mathrm{k}}^2\gtrsim m^2(1+\epsilon) $ which is bounded by
$ O(1) $ because
\[
\frac{\hat{\mathrm{k}}}{\hat{\mathrm{k}}^{2}}\cdot
\nabla_{\hat{\mathrm{k}}}\,\chi(\,\hat{\mathrm{k}}\,)=
-\;\frac{\zeta}{\sqrt{\hat{\mathrm{k}}^2}}\cdot
\frac{\hat{\rm k}^2 + m^2}
{(\,\hat{\rm k}^2 - m^2\,)^{2}}<0;
\qquad
\chi\left(\,\hat{\mathrm{k}}=m\sqrt{1+\epsilon}\,\right)=
\frac{\zeta\sqrt{1+\epsilon}}{m\epsilon}=1+\frac{\epsilon}{2}+\ldots
\]
Then the Bogolyubov transmission coefficient can be recast under the suitable form
\begin{eqnarray*}
\alpha(\,\hat{\mathrm{k}}\,)_{\pm}\,\equiv\,{\textstyle\frac12}\cdot\frac{K_{\pm}+K}{\surd(KK_{\pm})}=
\frac{1+\sqrt{\,1\pm\chi(\hat{\mathrm{k}})}}
{2\sqrt[4]{\,1\pm\chi(\hat{\mathrm{k}})}}
=\frac{1}{{2\sqrt[4]{\,1\pm\chi(\hat{\mathrm{k}})}}}+
{\textstyle\frac12}\,\sqrt[4]{\,1\pm\chi(\hat{\mathrm{k}})}\\
\end{eqnarray*}
while the integration volume in Fourier space is correspondingly bounded by
\[
\int\mathrm{d}\hat{\mathrm{k}}\equiv
\int_{-M}^{\,M}\mathrm d\omega\,\theta(\omega^2-m^2(1+\epsilon))
\int\!\!\!\!\int_{-\infty}^{\infty}\mathrm dk_y\,\mathrm dk_z\,
\theta(\,\omega^2 -m^2(1+\epsilon)- k_{x}^{2} - k_{y}^{2}\,)
\]
Then we can approximate by keeping the leading term
\begin{eqnarray*}
\ln\alpha(\,\hat{\mathrm{k}}\,)&=&\ln\left[\,1+\sqrt{\,1\pm\chi(\hat{\mathrm{k}})}\;\right]
-\ln2-{\textstyle\frac14}\,\ln\left[\,1\pm\chi(\hat{\mathrm{k}})\,\right]\\
&=&\pm{\textstyle\frac14}\,\chi(\hat{\mathrm{k}})
-{\textstyle\frac12}\cdot{\textstyle\frac18}\,\chi^{2}(\hat{\mathrm{k}})
- {\textstyle\frac{1}{32}}\,\chi^{2}(\hat{\mathrm{k}}) +\cdots
\mp{\textstyle\frac14}\,\chi(\hat{\mathrm{k}})+{\textstyle\frac18}\,\chi^{2}(\hat{\mathrm{k}})
+\cdots\\
&\approx& {\textstyle\frac{1}{32}}\,\chi^{2}(\hat{\mathrm{k}}) \qquad\quad
\mathrm{for}\quad \chi(\hat{\mathrm{k}})< 1
\end{eqnarray*}
within the phase space volume.
Thus we approximately obtain
\[
\int\mathrm{d}\hat{\mathrm{k}}\;\ln\alpha(\hat{\mathrm{k}})\approx
\frac{\zeta^2}{16}\int_{0}^{\,M}\mathrm d\omega\,\theta(\omega^2-m^2(1+\epsilon))
\int\!\!\!\!\int_{-\infty}^{\infty}\mathrm dk_y\mathrm dk_z
\theta(\hat{\mathrm{k}}^2 -m^2(1+\epsilon))
\frac{\hat{\mathrm{k}}^2}{(\hat{\rm k}^2 - m^2)^2}
\]
and if we set $ q=k_y^2+k_z^2 $ after turning to planar polar coordinates we obtain
\begin{eqnarray*}
\int\mathrm{d}\hat{\mathrm{k}}\ln\alpha(\hat{\mathrm{k}})\approx
\frac{\pi\zeta^2}{16}\int_{0}^{M}\mathrm d\omega\theta(\omega^2-m^2(1+\epsilon))
\int_{0}^{\infty}\mathrm{d}q \theta(\omega^2-m^2(1+\epsilon)-q)
\frac{\omega^2-q}{(\omega^2-m^2-q)^2}
\end{eqnarray*}
To evaluate the above integral it is convenient to set
$$ \upsilon=\omega^2-m^2(1+\epsilon) \qquad a=\sqrt{1+\epsilon}\qquad b=\frac{M}{m} $$
in such a manner that we can write

\begin{eqnarray*}
\int\mathrm{d}\hat{\mathrm{k}}\; \ln\alpha(\hat{\mathrm{k}}) &\approx&
\frac{\pi\zeta^2}{16}\int_{0}^{\,M}\mathrm{d}\omega\,
\theta(\upsilon)\,\frac{\partial}{\partial\xi}
\int_{0}^{\,\upsilon}\mathrm{d}q\;\left.
\frac{\omega^2-q}{\upsilon-q + \epsilon\,m^{2}- \xi}\;\right\rfloor_{\xi=0}\\
&=&
\frac{\pi\zeta^2}{16}\int_{0}^{\,M}\mathrm d\omega\,
\theta(\upsilon)\;\frac{\partial}{\partial\,\xi}
\Big\{(m^{2} + \xi)\Big[\,
\ln(\upsilon + \epsilon\,m^{2} -\xi)-\ln(\epsilon\,m^{2} - \xi)\,\Big]\Big\}_{\xi=0}\\
&=&
\frac{\pi}{16}\;m^{3}\epsilon^{\,2}\int_{a}^{\,b}\mathrm{d}x\,
\left[\,\ln(1+\varkappa) + \frac{1}{\epsilon}\cdot\frac{\varkappa}{1+\varkappa}\,\right]
\end{eqnarray*}
where we have set
\[
\varkappa=\frac{\upsilon}{\epsilon\,m^{2}}=
\frac{\omega^{2}-m^{2}(1+\epsilon)}{\epsilon\,m^{2}}=
\frac{1}{\epsilon}\left(x^{2} - 1\right)- 1
\]
Then we find
\begin{eqnarray*}
\int\mathrm{d}\hat{\mathrm{k}}\;\ln\alpha(\hat{\mathrm{k}})\approx
\frac{\pi}{16}\;m^{3}\epsilon^{\,2}\int_{a}^{\,b}\mathrm{d}x\;\left[\,
\ln(x+1)+\ln(x-1)-\ln\epsilon +\frac{1}{\epsilon}  +
\frac12\left(\frac{1}{x+1}-\frac{1}{x-1}\right)\right]
\end{eqnarray*}
so that, taking the assumption $ 0\le\epsilon\ll1\ll b $ properly into account,
\begin{eqnarray*}
\int\mathrm{d}\hat{\mathrm{k}}\;\ln\alpha(\hat{\mathrm{k}})&\approx&
\frac{\pi b}{16}\;m^{3}\epsilon^{\,2}\left\lbrace \frac{1}{\epsilon}
+ 2\ln b - \ln\epsilon + O(1)\,\right\rbrace\\
&\approx&
\frac{\pi b}{16}\;m^{3}\epsilon\left\lbrace 1 + \epsilon\ln\frac{b^{2}}{\epsilon}
+ O(\epsilon)\right\rbrace
\end{eqnarray*}
where we have neglected also by terms of order $ \ln(\epsilon)/b  < O(1)$
Thereby, under the assumption $ M\gg m\gg \zeta $,
 we eventually find that, to the leading order in the relevant LIV numbers $ \epsilon\ll1 $ and
$ b\gg1 $, the dominant contribution to the vacuum to vacuum persistence probability is provided by the flux density
\[
\frac{1}{\Delta t\Delta\sigma}\,\ln |\,\langle\,\Omega\;{\rm in}\,|\,0\;{\rm out}\,\rangle\,|
\approx
-\,\zeta\;\frac{mMc^{\,3}}{16\pi^{2}\hslash^{2}}\left\lbrace
1 + \frac{\hslash\zeta}{2mc}\left( 2\ln\frac{M}{m} -
\ln\frac{\hslash\zeta}{2mc}\right) + O\left(\frac{\hslash\zeta}{mc}\right)\right\rbrace
\]
where we have taken into account a factor two, because both LIV polarization contribute
the same amount to the leading order in this approximation.

\end{document}